\documentclass[prd, preprint, nofootinbib]{revtex4}
\usepackage{graphicx}
\usepackage{amsfonts}
\usepackage{latexsym}
\usepackage{amssymb}
\usepackage{epsfig}
\usepackage{color}

\begin{document}

\title{Axion monodromy inflation with multi-natural modulations 
}

\author{Tetsutaro Higaki, Tatsuo Kobayashi$^2$, Osamu Seto$^3$, and Yuya Yamaguchi$^2$}
 \affiliation{$^1$ Theory Center, KEK, 1-1 Oho, Tsukuba, Ibaraki 305-0801, Japan\\
$^2$Department of Physics, Hokkaido University, Sapporo 060-0810, Japan \\
$^3$Department of Life Science and Technology,
  Hokkai-Gakuen University, Sapporo 062-8605, Japan
}

%
\begin{abstract}
We study parameter space in the axion monodromy inflation corrected by dynamically generated terms 
involving with the axion.
The potential has the linear term with multiple sinusoidal functions, which play a role in generating modulations.
We show that this potential leads both to a large tensor-to-scalar ratio $r_T \sim 0.16$
and to a large negative running of spectral index $\alpha_s \sim - (0.02 -0.03)$,
ameliorating the tension between the result of the PLANCK and that of the BICEP2.
To realize these results,
a small hierarchy among dynamical scales is required whereas
the decay constants in sinusoidal functions remain sub-Planckian in this model.
We discuss also reheating process after the inflation in a bottom-up approach.
%
\end{abstract}

\pacs{}
\preprint{KEK-TH-1732}
\preprint{EPHOU-14-010}
\preprint{HGU-CAP-034}

\vspace*{3cm}
\maketitle


\section{Introduction}

Recently, BICEP2~\cite{Ade:2014xna} announced the detection of 
the rather large primordial B-mode fluctuation in the cosmic microwave background (CMB):
The tensor-to-scalar ratio is given by
\begin{equation}\label{eq:BICEP2-1}
r _T = 0.20^{+0.07}_{-0.05} ,
\end{equation}
for a lensed-$\Lambda$CDM plus tensor mode cosmological model.
It is estimated that
%
$r _T = 0.16^{+0.06}_{-0.05}$ 
after subtracting the foreground based on dust models.
The discovery of the tensor mode provides us information about the inflation in the early universe.
Such a large $r_T$ suggests 
that a large field inflation scenario such as chaotic inflation models \cite{Linde:1983gd,ChaoticAfterBicep} 
where the field value is super-Planckian
\cite{Lyth:1996im}.
Then the energy scale of the inflation is so high:
\begin{equation}
V^{1/4} \simeq 2.0 \times 10^{16}\,{\rm GeV} \cdot \left(\frac{r_T}{0.16} \right)^{1/4},
\end{equation} 
combined with the data from the PLANCK \cite{Ade:2013zuv}, where $V$ is the potential energy during the inflation.
The PLANCK observes also the spectral index $n_s$:
\begin{equation}
n_s = 0.9603 \pm 0.0073.
\end{equation} 
However, there exists a tension between PLANCK \cite{Ade:2013zuv} and BICEP2 
on the relative size of scalar density perturbations on large and small scales in the CMB.
Such a tension can be ameliorated when a negative running of the spectral index, $\alpha_s$, 
is taken into account.
It is shown that $\alpha_s = - (0.02 -0.03)$ works for it \cite{Ade:2013zuv,Ade:2014xna}
(see also \cite{Li:2014cka}).
Alternatively, the tension can be relaxed also if one considers an extra relativistic component \cite{Cicoli:2012aq,Dvorkin:2014lea}, 
non-zero neutrino mass~\cite{Ade:2014xna,Giusarma:2014zza,Dvorkin:2014lea}, an anti-correlation between
tensor and scalar modes~\cite{Contaldi:2014zua} or between tensor and
isocurvature modes~\cite{Kawasaki:2014lqa}.\footnote{ 
See also
\cite{Miranda:2014wga,Freivogel:2014hca,Higaki:2014pja} for other solutions.}
To this end, it is worth 
studying a single model in which one obtains
a large $r_T$, a negative large $\alpha_s$ in addition 
to a spectral index $n_s \sim 0.96$ and an enough e-folding number $N \geq 50$.

{}To construct such a model in the field theory,
the potential during the inflation needs to be under a good control over a super-Planckian value.
An axion $\phi$ is a good candidate for the inflaton 
owing to a shift symmetry:
\begin{eqnarray}
\phi \to \phi + C,
\end{eqnarray} 
where $C$ is a real parameter
 and this shift symmetry is an (indirect) consequence of the periodicity of the axion. 
In this sense, the natural inflation is an interesting possibility 
\cite{Freese:1990rb,Freese:2014nla,NaturalAfterBICEP2,Czerny:2014wza,Czerny:2014wua,Higaki:2014pja},
because this is a large field inflation model.
However, it is required that a super-Planckian value of the axion decay constant \cite{Ade:2013zuv,Freese:2014nla}
or an alignment mechanism with sub-Planckian decay constants of multiple axions 
\cite{Kim:2004rp,Czerny:2014wza,Higaki:2014pja}.\footnote{
See also \cite{Dimopoulos:2005ac}.
}
Then, we might have to worry about a control of the theory or a tuning among the decay constants.

On the other hand, the monodromy inflation 
\cite{Silverstein:2008sg,McAllister:2008hb,Flauger:2009ab,Dong:2010in}
is also inflation model with an axion,
where the linear potential 
can be generated for the axion (see also \cite{Kaloper:2008fb}):
\footnote{See for a review \cite{Baumann:2014nda}.}
\begin{eqnarray}
V_{\rm monodromy}(\phi) \propto  \phi.
\end{eqnarray} 
This is also a large field inflation model based on the chaotic inflation.
The apparent but small violation of the axionic shift symmetry, i.e., monodromy, is produced by the presence
of NS5-branes in a warped geometry 
in the extra dimension. 
The fundamental periodicity of the axion would be of ${\cal O}(M_{\rm Pl})$ or smaller,
 but the monodromy makes the axion unwrap just one cycle and spiral.
The field value is small and gravitational corrections are suppressed,
 but effectively the axion excursion during the inflation is long
 and it behaviors the large field inflation.
The usage of the NS5-branes protect the slow-roll condition against supergravity corrections
and generates the linear potential at a large value of the axion field. 
(See Ref.\cite{Marchesano:2014mla} also for recent discussion on the Wilson line monodromy inflation.)
The linear potential does not lead to a large negative running of 
spectral index and also slightly smaller tensor-to-scalar ratio, $r_T \sim 0.08$ is just 
realized \cite{Harigaya:2014sua,Kobayashi:2014ooa}.
In this model, no super-Planckian decay constants are required.

There would generally exist non-perturbative dynamics such as gaugino condensations and instantons,
which would be induced by the presence of the branes in the string theory \cite{Dine:1985rz,Witten:1996bn,Grimm:2011dj}.
Such effects should be naturally included because gauge theories including the Standard Model are living on the brane;
multiple sinusoidal corrections in the axion potential can be generated:
\begin{eqnarray}
V_{\rm modulation}(\phi) = \sum_i \Lambda_i^4 \cos \left(\frac{\phi}{f_i} + \delta_i \right) .
\end{eqnarray} 
Here
$\Lambda_i$ are dynamically generated scales and exponentially suppressed compared to the Planck scale.
$f_i$ are the decay constants in the respective terms and $\delta_i$ are phases.
This type of potential is discussed also for the multi-natural inflation in cases that
either decay constant takes a super-Planckian value \cite{Czerny:2014wza,Czerny:2014wua,Higaki:2014pja}. 
Even if these corrections are subdominant during the inflaton, 
they are important for observable parameters such as 
the spectral index $n_s$, its running $\alpha_s$ and tensor-to-scalar ratio $r_T$.
This is because large contributions to them appear
when differentiating the potential \cite{Flauger:2009ab,Kobayashi:2010pz,Czerny:2014wua,Kobayashi:2014ooa}.
Indeed, in Ref.\cite{Kobayashi:2014ooa} three of the present authors (T.K., O.S. and Y.Y.) have shown it is possible
to realize a large tensor mode, $r_T\sim 0.16$, and $n_s \sim 0.96$.
However, it is impossible to simultaneously obtain a large negative running, $\alpha_s \sim -0.02$, and a large $N \geq 50$
with the linear potential corrected by a single sinusoidal function.
This is because of the special relation between them (see Eq.(\ref{Eq:r-alpha})).
If $N$ is not large enough, it would be acceptable in the double inflation scenario \cite{DoubleInflation}:
This axion inflation is the first stage of the double inflation, 
and an another inflation follows to give an enough e-folding number, e.g., $N \geq 40$.\footnote{
Such a possibility for the double inflation was studied \cite{Kobayashi:2014rla,Choi:2014aca}.}

In this paper, we study parameter space in the axion monodromy inflation.
The potential consists of the linear potential and two sinusoidal terms, 
which are assumed to be generated by non-perturbative effects, for simplicity.
Then, we will see that such a potential can lead to quantities consistent with observations:  
$n_s \sim 0.96$, $r_T \sim 0.16$, $\alpha_s \sim -(0.02-0.03)$ and 
$N \geq 50$ at the same time in the single inflation scenario.
In special, larger spectral index around unity would be allowed in the presence of
extra relativistic freedom $N_{\rm eff} > 3$,
when we take the tension mentioned above into account \cite{Dvorkin:2014lea}.
Thus, we will consider also rather large $n_s$, and
also discuss parameter dependence of these observables.

This paper is organized as follows.
In section II, we explain explicitly our model.
In section III, we study observables in our model.
Section IV is devoted to the conclusion and discussion.

\section{Model}

In this section, we introduce the model with the linear potential and two sinusoidal functions.

\subsection{Review of slow-roll parameters}

Here, we review the convention of slow-roll parameters.
Here and hereafter, we use the Planck unit: $M_{\rm Pl}=2.4\times 10^{18}\,{\rm GeV} =1$.
The slow-roll parameters of 
the potential $V(\phi)$ of the inflaton $\phi$ are defined as follows,
\begin{eqnarray}
 \varepsilon &=& \frac{1}{2}\left(\frac{V_{\phi}}{V}\right)^2 , \\
 \eta &=& \frac{V_{\phi\phi}}{V} , \\
 \xi &=& \frac{V_{\phi}V_{\phi\phi\phi}}{V^2} ,
\end{eqnarray}
where $V_\phi$, $V_{\phi \phi}$ and $V_{\phi \phi \phi}$ 
denote the first, second and third derivatives of $V(\phi)$, 
respectively.

The scale of the scalar perturbation is written by 
\begin{eqnarray}
{\cal P}_{\zeta} &=& \left(\frac{H^2}{2\pi |\dot{\phi}|}\right)^2
 = \frac{V}{24 \pi^2 \varepsilon}.
 \end{eqnarray}
Here dot represents the first derivative with respect to the time.
The spectral index, its running and the tensor-to-scalar ratio are expressed as
\begin{eqnarray}
n_s &=& 1 + 2 \eta -6 \varepsilon ,\label{Formula:ns}\\
\alpha_s &=& 16 \varepsilon\eta -24 \varepsilon^2 -2\xi ,\label{Formula:alphas}\\
r_T &=& 16 \varepsilon,
\label{Formula:rT}
\end{eqnarray}
 at the first order of slow-roll parameters.
We note that the second order correction of $n_s$ will be included when those are important later (see Eq.(\ref{ns_high2})).
In addition, the number of e-folding is evaluated by the equation
\begin{eqnarray}
N = \int^{\phi}_{\phi_e} \frac{V}{V_{\phi}} d\phi.
\end{eqnarray}
Here, $\phi_e$ denotes the field value 
at the end of inflation, where the slow-roll condition is not satisfied, i.e., $\varepsilon~{\rm or}~ |\eta| \sim 1$.

\subsection{Potential}

We will study the inflation potential with two modulating terms for simplicity \cite{McAllister:2008hb,Flauger:2009ab}:
\begin{eqnarray}
\nonumber
V & = & V_{\rm monodromy} (\phi) + V_{\rm modulations} (\phi) \\
& = & a_1 \phi + a_2 \cos \left( \frac{\phi}{f} + \delta \right) + 
a'_2 \cos \left( \frac{\phi}{f'} + \delta' \right) + v_0, 
\label{potential}
\end{eqnarray}
where $\phi$ is the inflaton and the constant $v_0$ is added such that $V=0$ at the potential minimum but we can neglect it in our case.\footnote{
We can absorb either $\delta$ or $\delta '$ into the definitions of $\phi$ and $v_0$. 
However this would not change our result significantly.
}
We will take 
\begin{eqnarray}
a_{1},~a_2,~a_2' \ll 1;~~~
f,~f' \lesssim 1; ~~~
0 \leq \delta < 2\pi,~
0 \leq \delta' < 2\pi,
\end{eqnarray}
where the condition that $a_1 \phi \gg a_2~{\rm or}~a_2'$ should be satisfied during the monodromy inflation.
So, it looks that the last three terms in $V$ can be neglected naively.
However, this is not the case, because they give significant contributions
to the slow-roll parameters when one differentiates this potential.

In the remaining part of this section, we will briefly explain microscopic description of this potential.
The inflaton $\phi$ originates from the RR two-form potential $C_2$ in type IIB orientifold. 
The first term corresponds to the potential of the axion monodromy inflation.
This appears in a large $\phi$ limit from the term $\sqrt{1+\phi^2}$, 
which is generated
by the tension of NS5-branes localized on a warped geometry in the extra dimension.\footnote{
The same number of anti-NS5-branes are also introduced to cancel charges on a distant but homologous two-cycle.
See Ref.\cite{Conlon:2011qp} for a discussion.
} 
The tension of the brane can become much smaller than the Planck scale
because of the warping factor;
$a_1 \ll 1$ is related with a redshifted tension of the brane.
Such NS5-branes are considered in order to evade the eta problem from supergravity corrections.
Note that
the brane breaks the axionic shift (gauge) symmetry of $\phi$;
this is called monodromy.
This term can produce the mass of the inflaton after the inflation.
As mentioned in Introduction, the fundamental periodicity of the axion would not be super-Planckian,
 but the axion does not wrap just one cycle and spiral.
That makes other violations of the shift symmetry small enough. 
The second and third terms can be generated by 
non-perturbative effects such as D-brane instantons or gaugino condensations near the NS5-branes; $a_2,~a_2' \ll 1$ are expected.
They will be generated on fluxed D-branes 
wrapping on the rigid divisors containing a two cycle which is odd under the orientifold parity \cite{Grimm:2011dj}.
Further, as $\phi$ comes from $C_2$, decay constants are expected to be around the winding scale:
$f \sim f' \sim {\cal V}^{-1/3} \ll 1$, where ${\cal V}$ is the volume of the extra dimension \cite{Grimm:2004uq,McAllister:2008hb}, and $f$ and $f'$ depend on the origin of such non-perturbative effect.
We assume that non-perturbative dynamics to generate these terms are independent of each other, 
i.e. D-brane instantons on two independent cycles or gaugino condensations of two independent gauge theories.
Thus, the parameters $f$ and $f'$ ($a_2$ and $a'_2$) are independent of each other.\footnote{
In examples in the next section, multi instanton effects are subdominant compared with these terms.}
The phase parameters would be affected by heavy moduli stabilized at UV scales.  
Non-perturbative gravitational effects such as wormholes might induce 
another correction, but that is beyond our scope.
Here we assume that such a correction is negligible.

In the next section, we will study effects of these parameters on the observables.

\section{Results}

\subsection{Simple analyses and requirement}

Before considering the full potential (\ref{potential}),
 let us review more simple situations.
First, we consider the limit with both $a_2 \rightarrow 0$ and $a_2' \rightarrow 0$,
 in which the results are quite simple.
We obtain 
\begin{eqnarray}
	\varepsilon = \frac{1}{2\phi^2}, \qquad \eta = \xi = 0,
\end{eqnarray}
and 
\begin{eqnarray}
	N = \frac{1}{2} (\phi^2 - \phi_e^2).
\end{eqnarray}
To solve the horizon and flatness problem, $N$ should be more than 50.
This is realized by $\phi \sim 10$, and then we estimate 
\begin{eqnarray}
	N \sim 50, \qquad n_s \sim 0.97, \qquad r_T \sim 0.08, \qquad \alpha_s \sim -0.0006,
\end{eqnarray}
which are consistent with PLANCK and BICEP2 except for $\alpha_s$
(see e.g. Ref.\cite{Harigaya:2014sua}).
A large negative value of $\alpha_s$ is not realized in this case.
That is obvious because we can write 
\begin{eqnarray}
	n_s -1 = -\frac{3}{2N}, \qquad r_T = \frac{4}{N}, 
	\qquad \alpha_s = -\frac{3}{2 N^2}.
\end{eqnarray}
We would have a large negative $\alpha_s$ for smaller $N$.
However, the $n_s$ becomes too small.

Next, we consider the limit with $a_2 \rightarrow 0$,
 which was studied in a previous work \cite{Kobayashi:2014ooa}.
Then, the slow-roll parameters can be written as 
\begin{eqnarray}
	\varepsilon &=& \frac{1}{2\phi^2}\left[ \frac{1-x \sin \theta}{1+A \phi^{-1} (\cos \theta - \cos \delta)}\right]^2, \\
	\eta &=& - \frac{x^2}{A \phi} \left[ \frac{\cos \theta}{1 + A \phi^{-1} (\cos \theta-\cos \delta)} \right] , \\
	\xi &=& -\frac{x}{A} \sqrt{2\varepsilon}\, \eta \tan \theta,
\end{eqnarray}
where $A = a_2/a_1$, $\theta = \phi/f + \delta$ and $x= A/f$. 
In the followings, we will take $x >0$ without a loss of generality. 
We can see that $N$ is almost the same as the case of vanishing $a_2$
as long as the sinusoidal term is subdominant in the potential.
Thus, $N \sim 50$ is realized by $\phi \sim 10$.

For large $\phi$ (we assume $\phi=10$ in the following), we could approximate 
\begin{eqnarray}
	\varepsilon \approx \frac{\left( {1-x \sin \theta}\right)^2}{2\phi^2}, \qquad 
	\eta \approx - \frac{x^2 \cos \theta}{A \phi}, \qquad
	\xi \approx \frac{x^3 \sin \theta (1-x \sin \theta)}{A^2 \phi^2}.
\label{appro}
\end{eqnarray}
To realize $n_s \sim 0.96$ and $r_T \sim 0.16$, we find
$\varepsilon \sim 0.01$ and $\eta \sim 0.01$ from
Eqs.~(\ref{Formula:ns}) and (\ref{Formula:rT}).
Then, to realize $\alpha_s = - {\cal O}(0.01)$ additionally, we require $\xi = + {\cal O}(0.01)$,
that is, $\sin \theta$ must be positive at least.
However, when $\sin \theta > 0$, we obtain $\varepsilon \lesssim 0.005$, or $r_T \lesssim 0.08$.
In other words, with Eq. (\ref{appro}), $r_T$ is approximately given by
\begin{eqnarray}
	r_T \approx \frac{1}{N} \left( 1 + \sqrt{1+ 4 \alpha_s N f^2  }\, \right)^2.
	\label{Eq:r-alpha}
\end{eqnarray}
Here, we have used $N \approx \phi^2/2$, $\alpha_s \approx -2 \xi$ and $A/x = f$. 
Thus, we can see that large $r_T$ and large negative $\alpha_s$ can not be simultaneously realized
in the linear potential with one sinusoidal correction term.
Note that if $\phi$ is smaller than 10 they can be realized,
while second inflation is needed to obtain $N \gtrsim 50$ \cite{Kobayashi:2014ooa}.

Now, we consider the full potential (\ref{potential}) which has two sinusoidal correction terms.
In this case, we will see all $n_s$, $r_T$ and $\alpha_s$ can be realized as observably favored values.
To see the result, let us analyze what parameter regions are favored.
The slow-roll parameters can be written as
\begin{eqnarray}
	\varepsilon &=& \frac{1}{2 \phi^2} \left[ \frac{1 - x \sin\theta - x' \sin\theta'}{1 + A \phi^{-1} (\cos \theta - \cos \delta) + A' \phi^{-1} (\cos \theta' - \cos \delta')} \right]^2, \label{eps}\\
	\eta &=& - \frac{1}{\phi} \left[ \frac{x^2 A^{-1} \cos\theta + x'^2 A'^{-1} \cos\theta'}{1 + A \phi^{-1} (\cos \theta - \cos \delta) + A' \phi^{-1} (\cos \theta' - \cos \delta')} \right], \label{eta}\\
	\xi &=& - \sqrt{2 \varepsilon}\, \eta \left( \frac{x^3 A^{-2} \sin \theta + x'^3 A'^{-2} \sin \theta'}{x^2 A^{-1} \cos \theta + x'^2 A'^{-1} \cos \theta'} \right), \label{xi}
\end{eqnarray}
where $A' = a'_2/a_1$, $\theta' = \phi/f' + \delta'$ and $x'= A'/f'$.
Hereafter, we will take both $x$ and $x'$ to be positive without a loss of generality. 
Here, $\xi$ is also written by
\begin{eqnarray}
	\xi = \sqrt{2 \varepsilon}\, \left( \frac{x^3 A^{-2} \sin \theta + x'^3 A'^{-2} \sin \theta'}{V/a_1} \right).\label{xi2}
\end{eqnarray}
Since $\xi$ must be positive to realize large negative $\alpha_s$,
 the numerator must be positive;
\begin{eqnarray}
	\frac{x^3}{A^2} \sin \theta + \frac{x'^3}{A'^2} \sin \theta' > 0. \label{positive_xi}
\end{eqnarray}

When the sinusoidal terms are subdominant in the potential,
 $N \sim 50$ is realized by $\phi \sim 10$,
 and then we could approximate 
\begin{eqnarray}
	\varepsilon \approx \frac{\left( {1-x \sin \theta -x' \sin \theta'}\right)^2}{2\phi^2}, \qquad 
	\eta \approx - \frac{1}{\phi} \left( \frac{x^2}{A} \cos\theta +\frac{x'^2}{A'} \cos\theta' \right).
\end{eqnarray}
Since $\varepsilon \sim 0.01$ to realize $r_T \sim 0.16$,
 the value in parentheses of $\varepsilon$ should be large as $\sqrt{2}$.
Thus, $\sin \theta$ or $\sin \theta'$ must be negative at least.
Here, we assume $\sin \theta < 0$, $\sin \theta' > 0$ and
\begin{eqnarray}
	x\, |\sin \theta| > x' \sin \theta'.\label{theta}
\end{eqnarray}
Then, $\sin \theta'$-dependent term should be dominant in Eq. (\ref{positive_xi})
to realize a negative $\alpha_s$.
Thus, we require also
\begin{eqnarray}
	\frac{x}{A} < \frac{x'}{A'}.\label{xA}
\end{eqnarray}
Using Eqs. (\ref{theta}) and (\ref{xA}),
 $\varepsilon$ and $\xi$ are approximately written by
\begin{eqnarray}
	\varepsilon \approx \frac{\left( {1-x \sin \theta}\right)^2}{2\phi^2}, \qquad 
	\xi \approx -\frac{x'}{A'} \sqrt{2\varepsilon}\, \eta \tan \theta'.
\label{appro2}
\end{eqnarray}
In fact, to realize observably favored values of all $n_s$, $r_T$ and $\alpha_s$,
 we require simple relations as $\varepsilon \sim \eta \sim \xi \sim 0.01$.
Then, the parameters should satisfy two equations given by
\begin{eqnarray}
	&&1-x \sin \theta \simeq \sqrt{2}, \label{rt_condition}\\
	&&\frac{x'}{A'} |\tan \theta'| \simeq {\cal O}(10).\label{al_condition}
\end{eqnarray}
Since these equations are valid only when the parameters satisfy Eqs. (\ref{theta}) and (\ref{xA}),
we can find the inequalities (or hierarchy between the parameters) are so important.
Note that both the second term and third term in the potential (\ref{potential}) are required
to realize large $r_T$ and large negative $\alpha_s$, respectively.

Finally, we point out higher order terms of $n_s$ can not be neglected
in models which lead to a large $\xi$ as of ${\cal O}(0.01)$.
The spectral index $n_s$ with the higher order terms is given by \cite{Lyth:1998xn}\footnote{
See also Ref. \cite{Gao:2014yra}.}
\begin{eqnarray}
	n_s =  1 + 2 \eta - 6 \varepsilon
		+ 2 \left[ \frac{1}{3} \eta^2 + (8C-1) \varepsilon \eta - \left( \frac{5}{3}+12C \right) \varepsilon^2 - \left( C-\frac{1}{3} \right) \xi \right],
\label{ns_high}
\end{eqnarray}
where $C = -2 + \ln 2 + \gamma \simeq -0.73$, and $\gamma \simeq 0.577$ is the Euler-Mascheroni constant.
Since $\xi$ can be small as ${\cal O}(0.001)$ in most large field inflation models, 
 the higher order terms can be neglected.
However, since now we consider $\varepsilon \sim \eta \sim \xi \sim 0.01$,
 the last term of Eq. (\ref{ns_high}) can not be neglected.
Particularly, when large negative $\alpha_s$ is almost given by $\xi$,
 we can estimate
\begin{eqnarray}
	n_s \approx 1 + 2 \eta - 6 \varepsilon + 2 \xi \approx 1 + 2 \eta - 6 \varepsilon + |\alpha_s|.
\label{ns_high2}
\end{eqnarray}
This enhancement by $|\alpha_s|$ is important
 because $n_s$ can be observed in a great accuracy.
Thus, we calculate $n_s$ with including the $\xi$ term in our numerical analyses.

\subsection{Numerical analyses}

%
\begin{figure}[t]
  \begin{center}
      \begin{minipage}{0.45\hsize}
        \begin{center}
          \includegraphics[clip, width=\hsize]{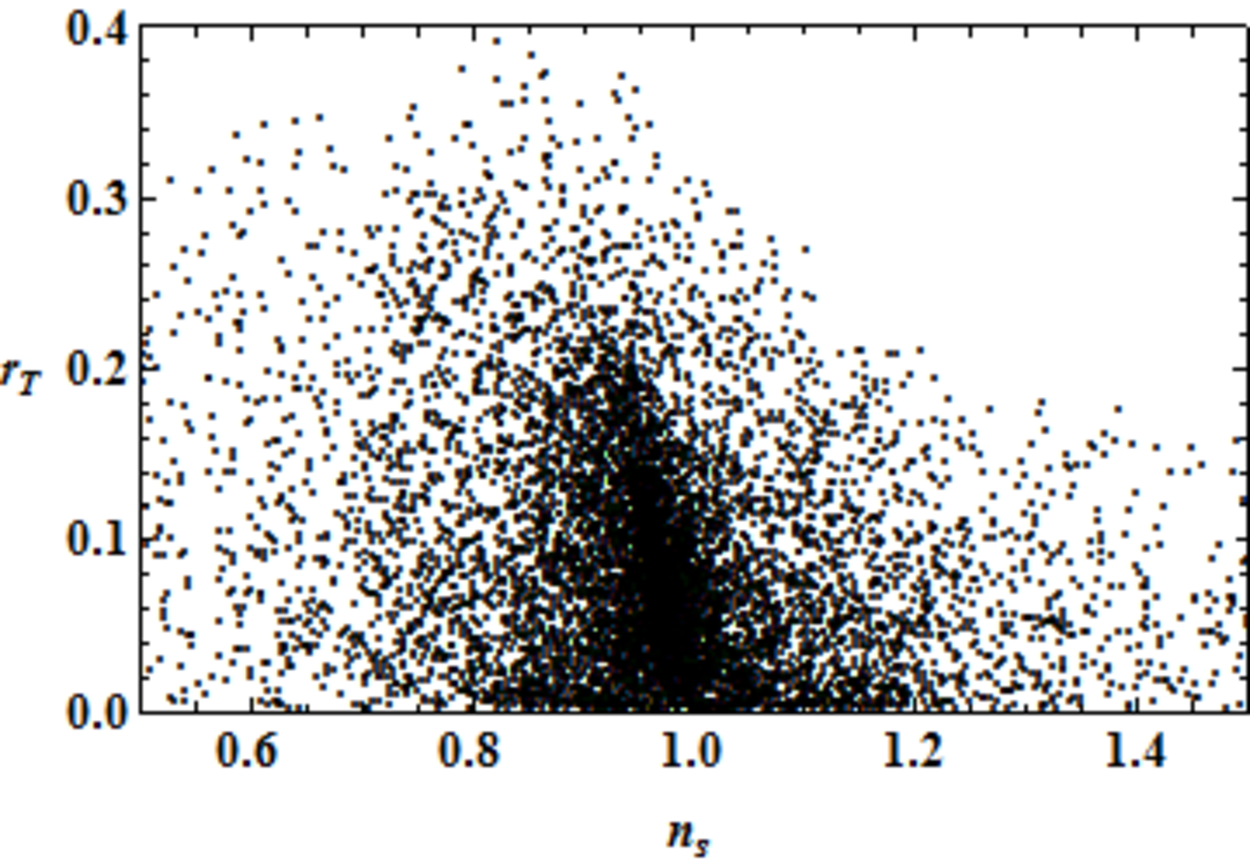}
        \end{center}
      \end{minipage}
\hspace{5mm}
      \begin{minipage}{0.45\hsize}
        \begin{center}
          \includegraphics[clip, width=\hsize]{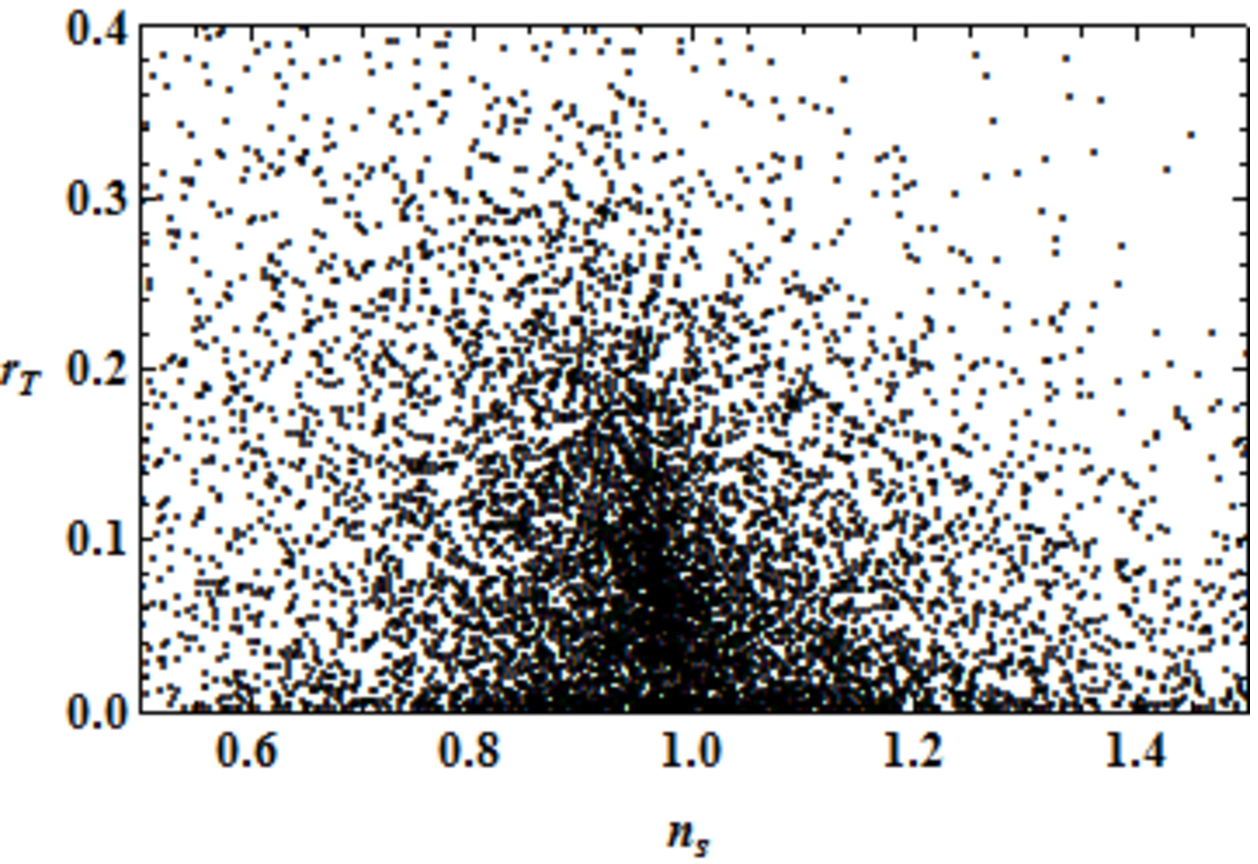}
        \end{center}
      \end{minipage}
  \end{center}
  \begin{center}
      \begin{minipage}{0.45\hsize}
        \begin{center}
          \includegraphics[clip, width=\hsize]{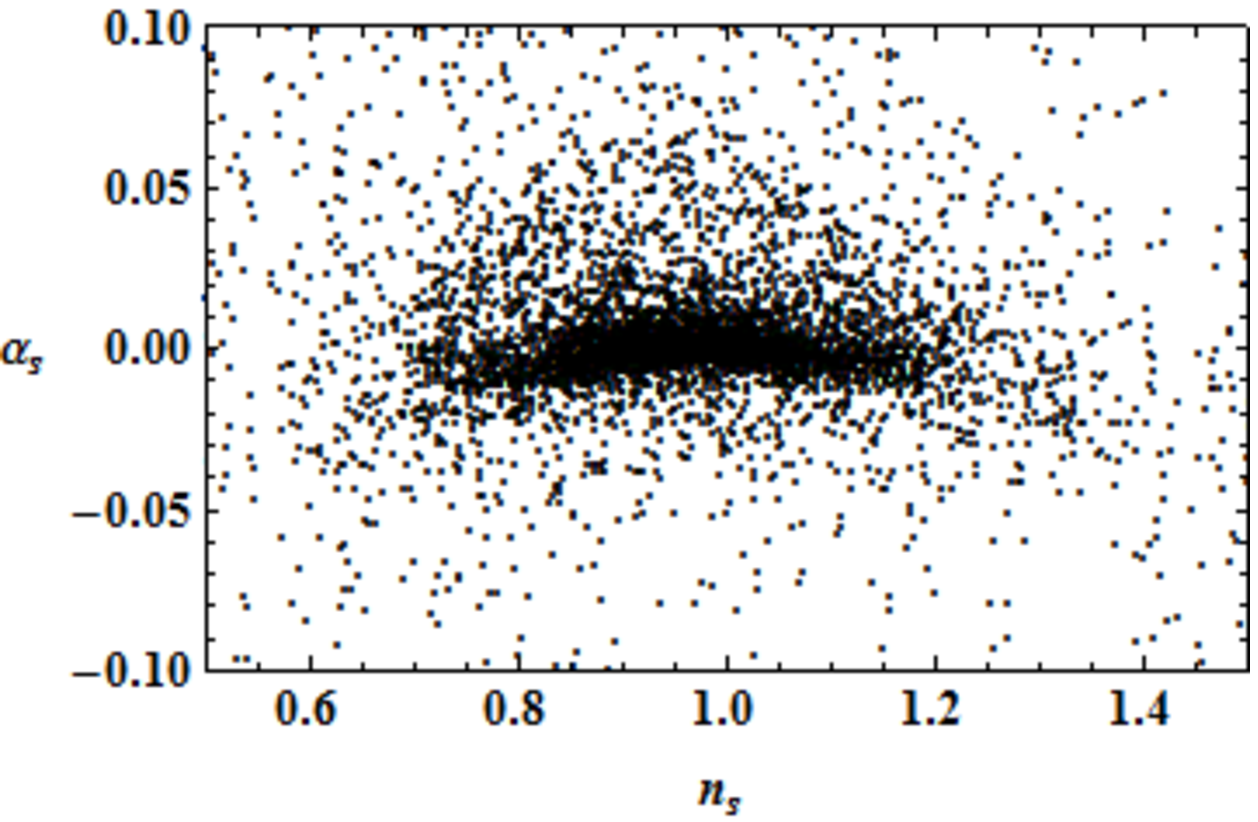}
        \end{center}
      \end{minipage}
\hspace{5mm}
      \begin{minipage}{0.45\hsize}
        \begin{center}
          \includegraphics[clip, width=\hsize]{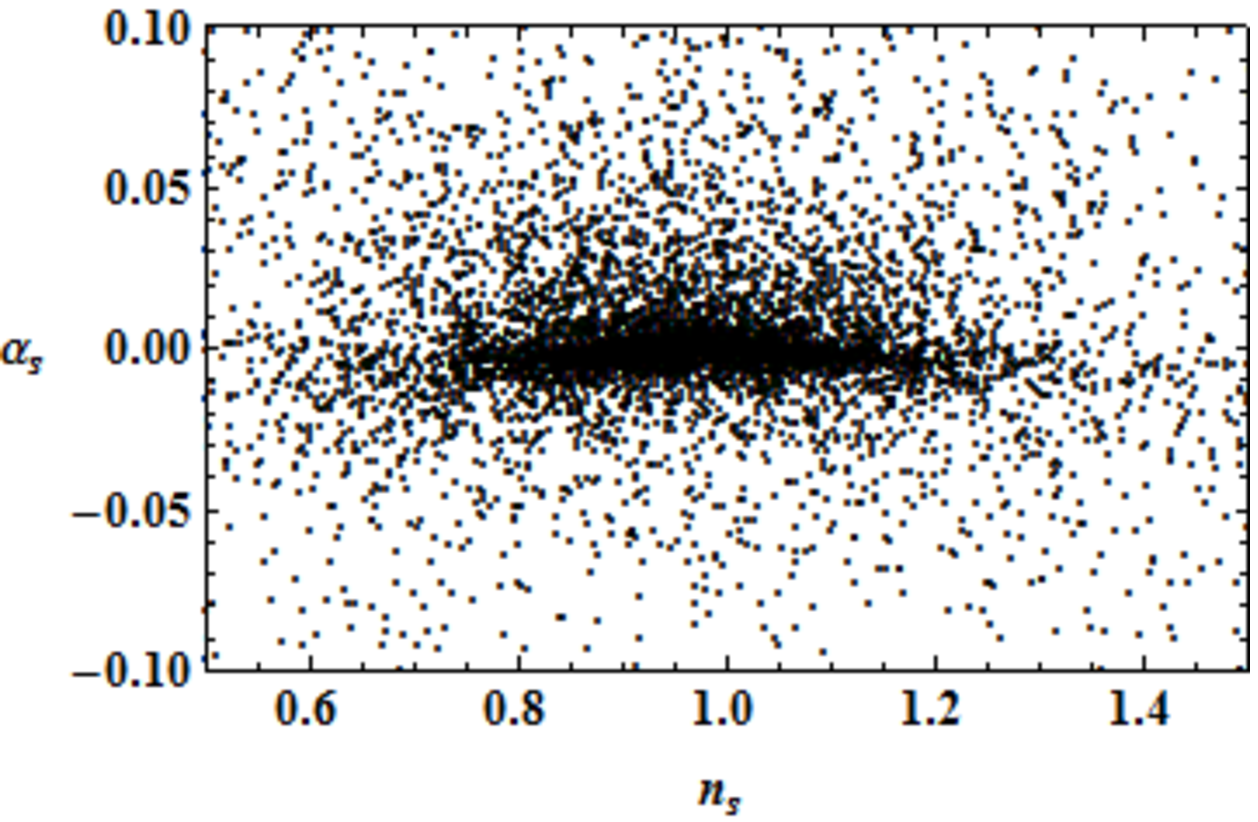}
        \end{center}
      \end{minipage}
  \end{center}
  \begin{center}
      \begin{minipage}{0.45\hsize}
        \begin{center}
          \includegraphics[clip, width=\hsize]{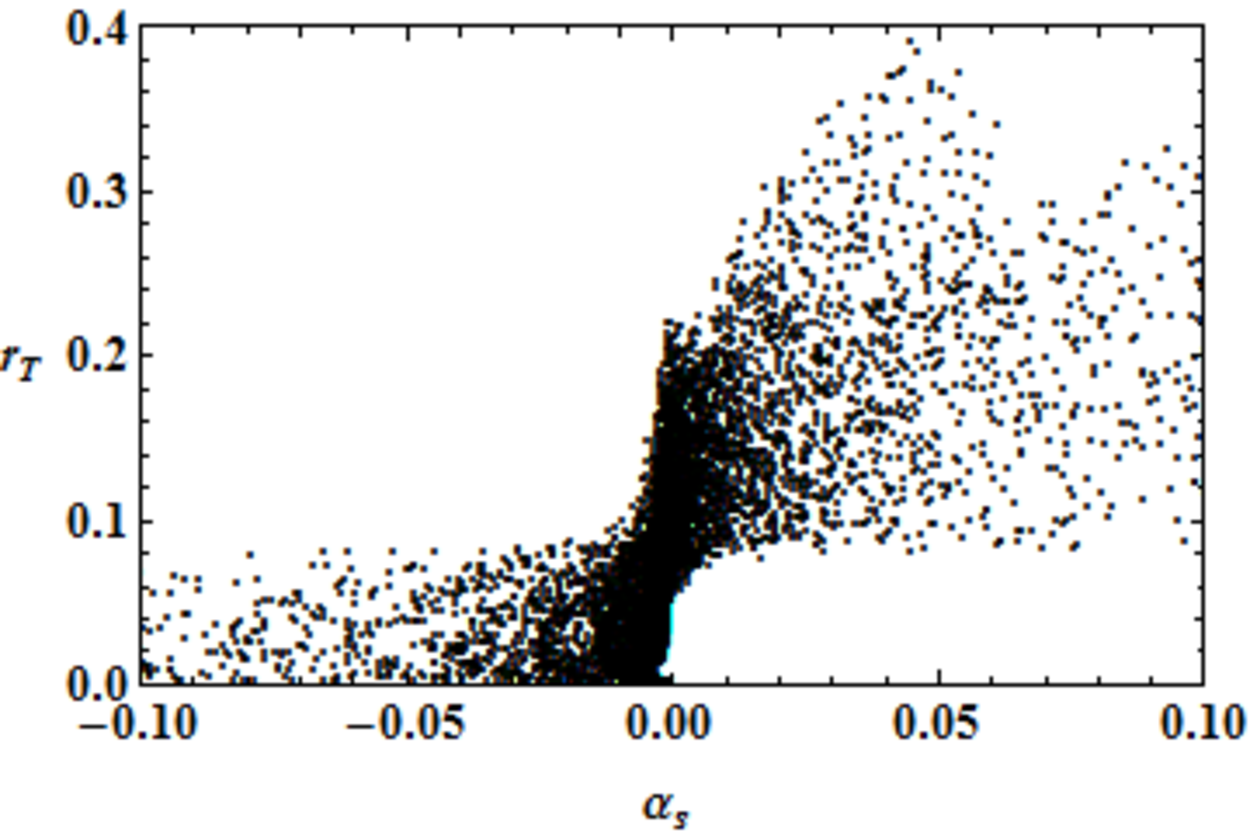}
        \end{center}
      \end{minipage}
\hspace{5mm}
      \begin{minipage}{0.45\hsize}
        \begin{center}
          \includegraphics[clip, width=\hsize]{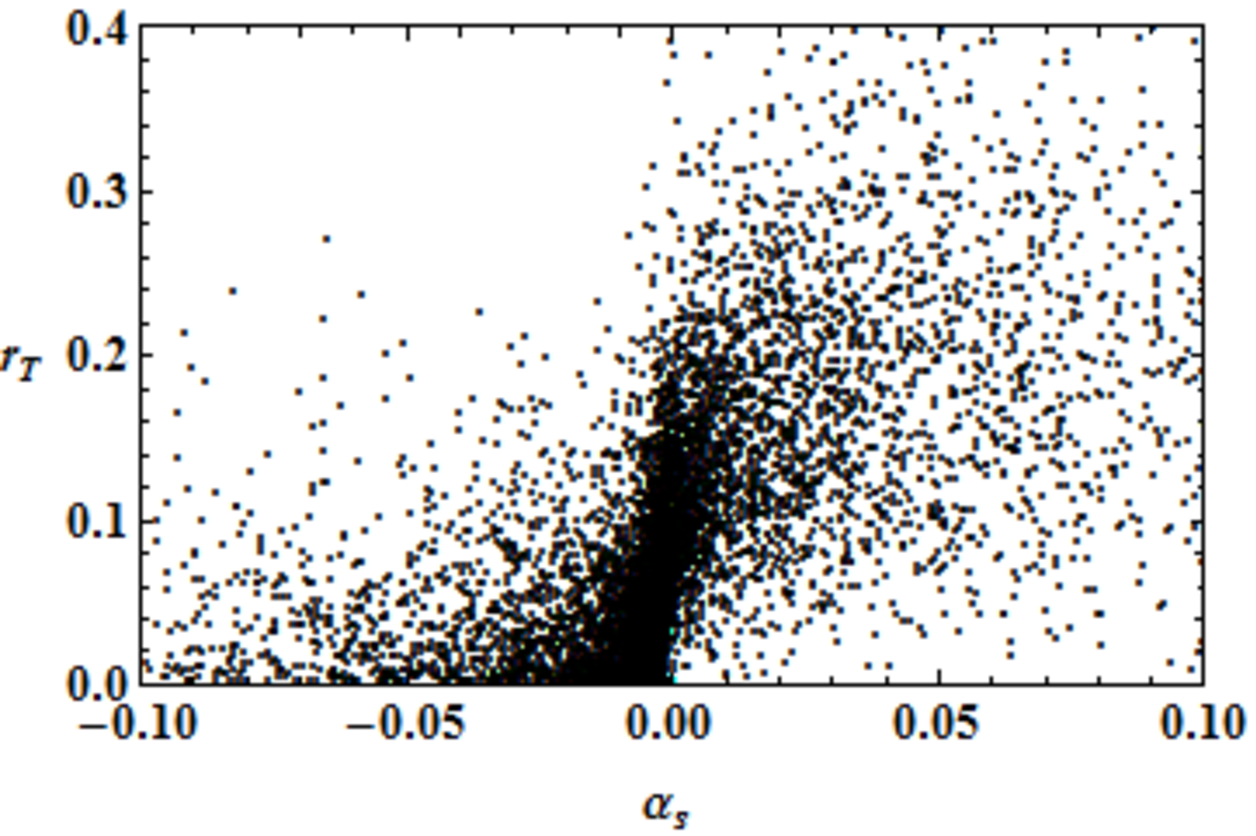}
        \end{center}
      \end{minipage}
  \end{center}
   \caption{
Scatter plots in $(n_s,~r_T)$, 
$(n_s,~\alpha_s)$ or $(\alpha_s,~r_T)$-planes for $a'_2 = 0$ (left) and for $a'_2 \neq 0$ (right). 
			Parameters are randomly chosen in the region as follows:
			$0.001 \leq x \leq 1$, $0.001 \leq x' \leq 1$, $0.001 \leq A \leq 1$, $0.001 \leq A' \leq 1$, $0 \leq \theta \leq 2 \pi$, and $0 \leq \theta' \leq 2 \pi$.}
	\label{cos_dif}
\end{figure}
%
We show the numerical results of above discussions.
Here, we assume $\phi = 10$ to obtain $N \sim 50$.
First, to see the general results of the potential with the sinusoidal corrections,
we show the scatter plots in Fig.\,\ref{cos_dif}, in which parameters are randomly chosen.
We can find that $n_s$ is uniformly distributed around the central value at $n_s \simeq1$,
 while $r_T$ and $\alpha_s$ are seemed to have some relation, which follows from Eq.(\ref{Eq:r-alpha}).
Eq.(\ref{Eq:r-alpha}) tells the reason why large $r_T$ and large negative $\alpha_s$ can not be simultaneously realized
for vanishing $a'_2$ as shown in Fig.\,\ref{cos_dif}.
Thus, when $\phi=10$, we obtain $r_T \lesssim 0.08$ for negative $\alpha_s$, while $r_T \gtrsim 0.08$ for positive $\alpha_s$.

In the followings. we shall consider cases with $a_2' \neq 0$.
Note that, in the this case,
all observable values can be distributed more widely than the cases for $a'_2=0$.
Particularly, large $r_T$ and large negative $\alpha_s$ can be simultaneously realized.

%
\begin{figure}[t]
  \begin{center}
      \begin{minipage}{0.31\hsize}
        \begin{center}
          \includegraphics[clip, width=\hsize]{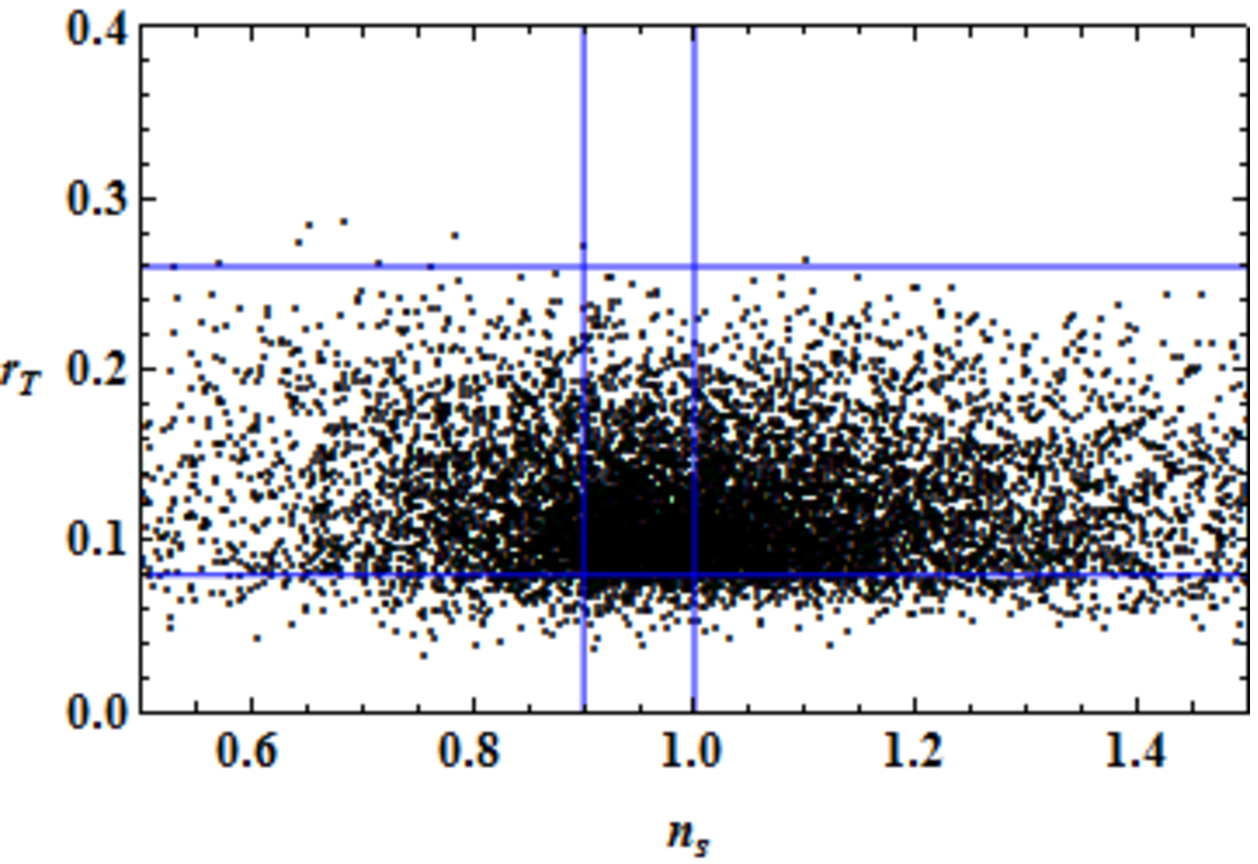}
        \end{center}
      \end{minipage}
\hspace{3mm}
      \begin{minipage}{0.31\hsize}
        \begin{center}
          \includegraphics[clip, width=\hsize]{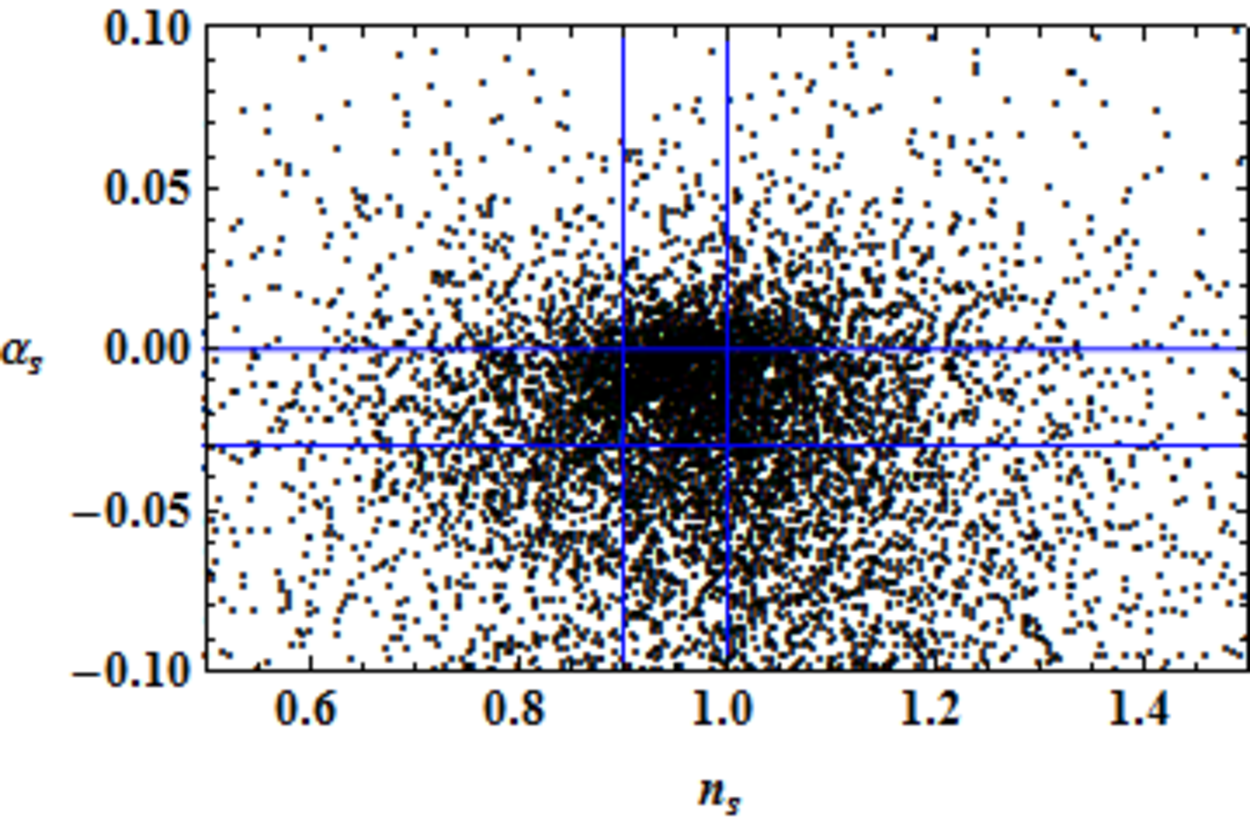}
        \end{center}
      \end{minipage}
\hspace{3mm}
      \begin{minipage}{0.31\hsize}
        \begin{center}
          \includegraphics[clip, width=\hsize]{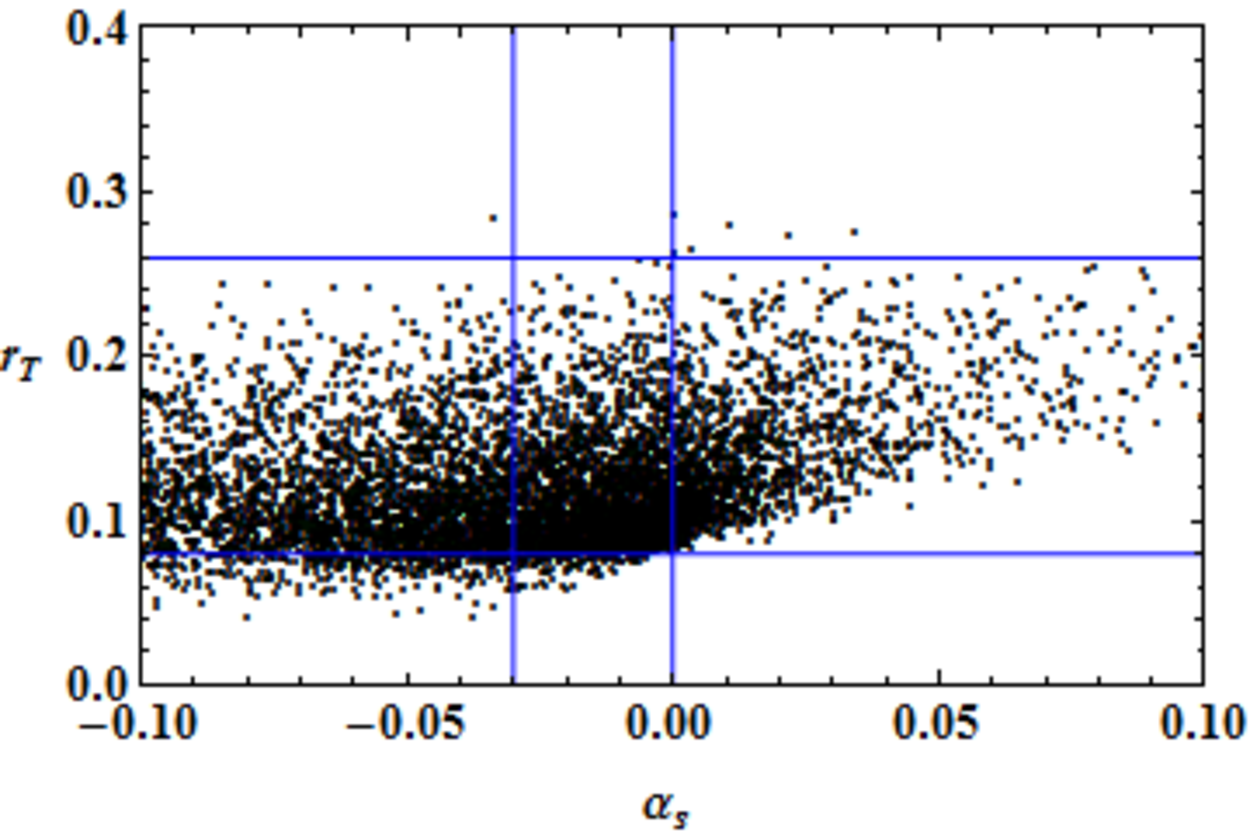}
        \end{center}
      \end{minipage}
  \end{center}
  \caption{
Scatter plots in $(n_s,~r_T)$, 
$(n_s,~\alpha_s)$ or $(\alpha_s,~r_T)$-planes for $a'_2 \neq 0$. 
The regions surrounded by blue lines correspond to the range such that
$0.9 \leq n_s \leq 1.0$, $0.08 \leq r_T \leq 0.26$ and $-0.03 \leq \alpha_s \leq 0$.
			Parameters are randomly chosen in the region as follows:
			$0.1 \leq x \leq 0.8$, $x/10 \leq x' \leq x/2$, $1 \leq x/A \leq 3$, $2 x/A \leq x'/A' \leq 5 x/A$, $\pi \leq \theta \leq 2 \pi$, and $0 \leq \theta' \leq \pi$.
}
\label{restrict}
\end{figure}
%
%
%
%
\begin{figure}[t]
  \begin{center}
      \begin{minipage}{0.45\hsize}
        \begin{center}
          \includegraphics[clip, width=\hsize]{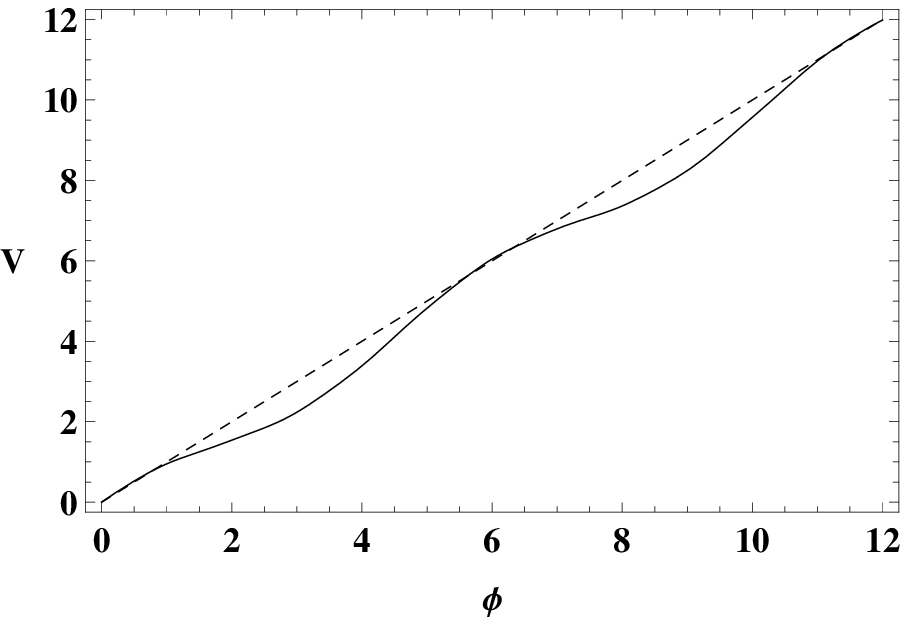}
        \end{center}
      \end{minipage}
\hspace{5mm}
      \begin{minipage}{0.45\hsize}
        \begin{center}
          \includegraphics[clip, width=\hsize]{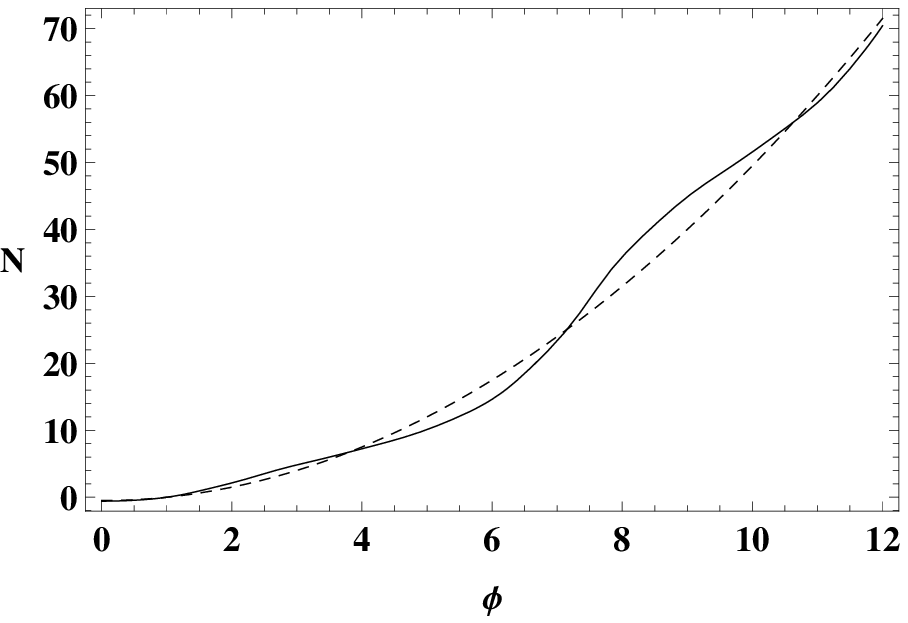}
        \end{center}
      \end{minipage}
  \end{center}
  \caption{
Plots of potential  (left) and of the number of e-fold (right)
versus inflaton value. In this figure, we used parameters in the Table 1. 
			The solid and dashed lines correspond to the potential given by Eq. (\ref{potential}) and with only linear term, respectively.}
  \label{example_fig1}
\end{figure}
%
\begin{figure}[!ht]
  \begin{center}
      \begin{minipage}{0.32\hsize}
        \begin{center}
          \includegraphics[clip, width=\hsize]{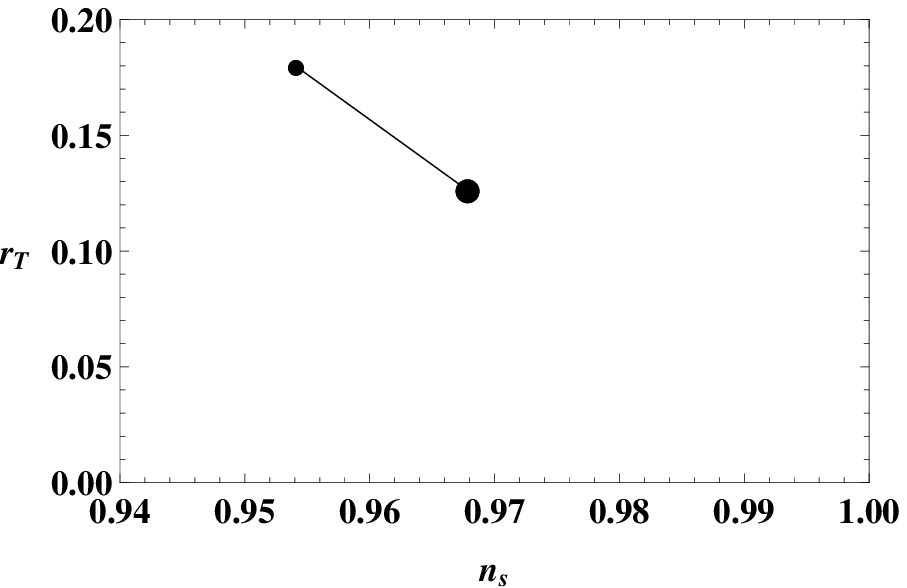}
        \end{center}
      \end{minipage}
\hspace{1mm}
      \begin{minipage}{0.32\hsize}
        \begin{center}
          \includegraphics[clip, width=\hsize]{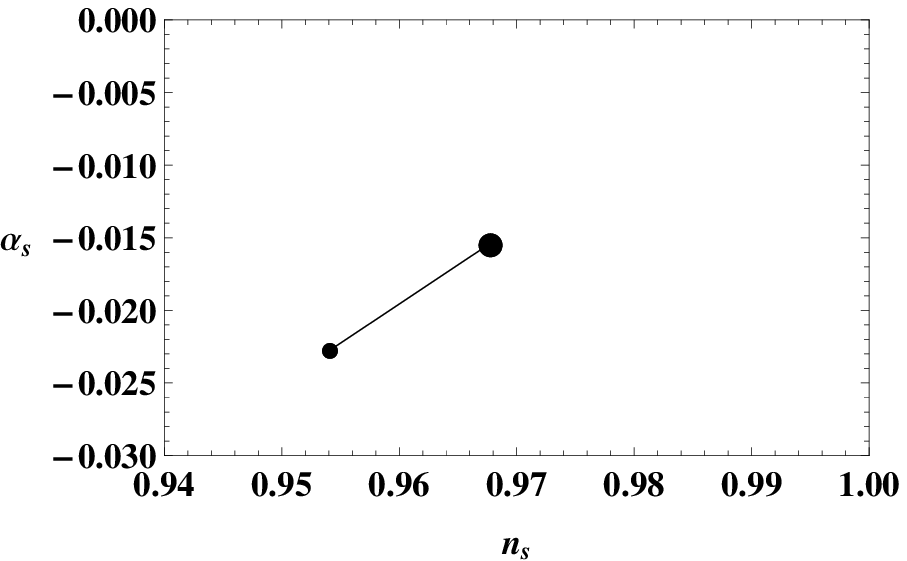}
        \end{center}
      \end{minipage}
\hspace{1mm}
      \begin{minipage}{0.32\hsize}
        \begin{center}
          \includegraphics[clip, width=\hsize]{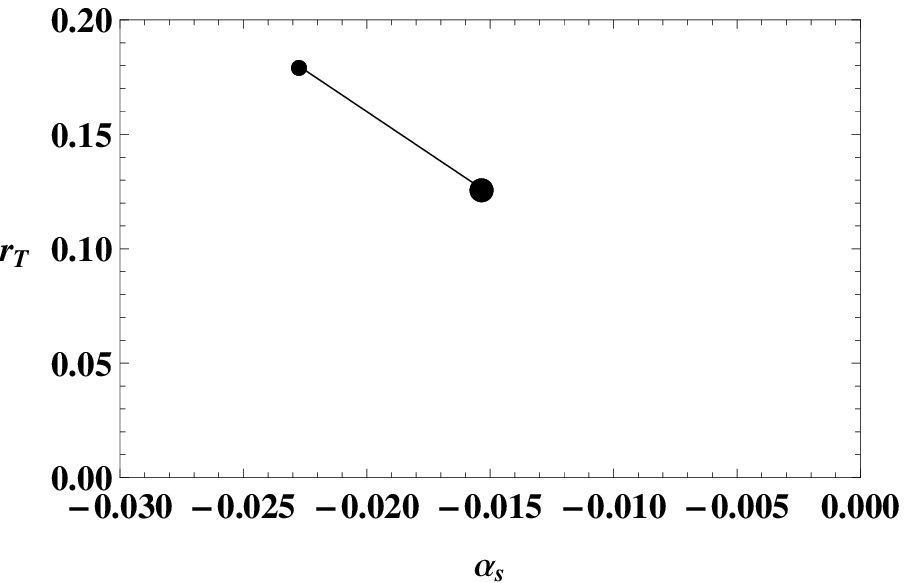}
        \end{center}
      \end{minipage}
  \end{center}
  \caption{
Plots of the number of e-fold in $(n_s,~r_T)$, 
$(n_s,~\alpha_s)$ or $(\alpha_s,~r_T)$-planes. 
We have used the parameters in Table \ref{example}. 
			Smaller and larger circle correspond to $N$ = 50 and 60 ($\phi \simeq 9.7$ and 11), respectively.}
  \label{example_fig2}
\end{figure}
%
\begin{table}[t]
\begin{center}
\begin{tabular}{|c|ccc|ccc|cccc|}\hline
~~~$\phi$~~~ & ~~~$x$~~~ & ~~~$A$~~~ & ~~~$\theta$~~~ & ~~~$x'$~~~ & ~~~$A'$~~~ & ~~~$\theta'$~~~ & ~~~$N$~~~ & ~~~$n_s$~~~ & ~~~$r_T$~~~ & ~~~$\alpha_s$~~~ \\
\hline \hline
10 & 0.45 & 0.4 & ~$260\, \pi/180$~ & 0.05 & 0.01 & ~$70\, \pi/180$~ & 52 & 0.956 & 0.17 & -0.021\\
\hline
\end{tabular}
\end{center}
\caption{A typical example which can realize the favored values of all $n_s$, $r_T$ and $\alpha_s$}
\label{example}
\end{table}
%
Fig.\,\ref{restrict} shows the scatter plots,
in which we restrict parameter regions to those satisfying Eqs. (\ref{theta}) and (\ref{xA}).
We are focusing also on regions with $f' < f < 1$, which are sub-Planckian decay constants.

We can see that a large $r_T$ and a large negative $\alpha_s$ can be simultaneously realized in many cases.
The hierarchy between $x$ and $x'$ is important for realizing a large $r_T$,
whereas the hierarchy between $x/A$ and $x'/A$ is important for obtaining a large negative 
$\alpha_s$.
Particularly, when we restrict $x$ and $x'$ as $0.1 \leq x \leq 0.8$ and $x/10 \leq x' \leq x/2$,
Eq.(\ref{appro2}) becomes valid and leads to range of $0.08 \lesssim r_T \lesssim 0.26$.
In the same way, when we restrict $x/A$ and $x'/A'$ as $1 \leq x/A \leq 3$ and $2 x/A \leq x'/A' \leq 5 x/A$,
Eq. (\ref{positive_xi}) is satisfied in most cases; $\alpha_s$ tends to be negative then.

In Fig.\,\ref{example_fig1} and Fig.\,\ref{example_fig2},
we show also typical examples,
in which we obtain favored values
of all $n_s$, $r_T$ and $\alpha_s$.
In these figures, we have used parameters shown in Table \ref{example}.
%
%
\begin{figure}[t]
  \begin{center}
      \begin{minipage}{0.32\hsize}
        \begin{center}
          \includegraphics[clip, width=\hsize]{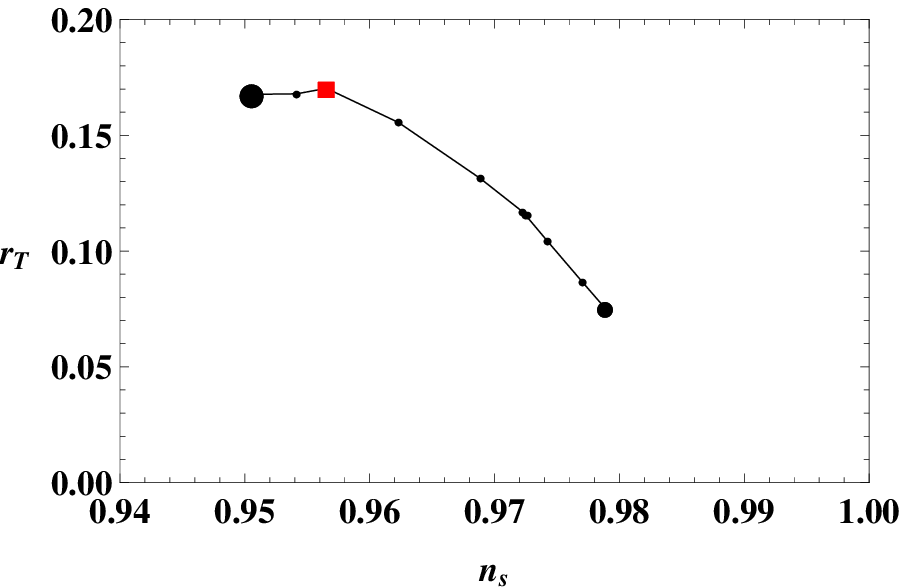}
        \end{center}
      \end{minipage}
\hspace{1mm}
      \begin{minipage}{0.32\hsize}
        \begin{center}
          \includegraphics[clip, width=\hsize]{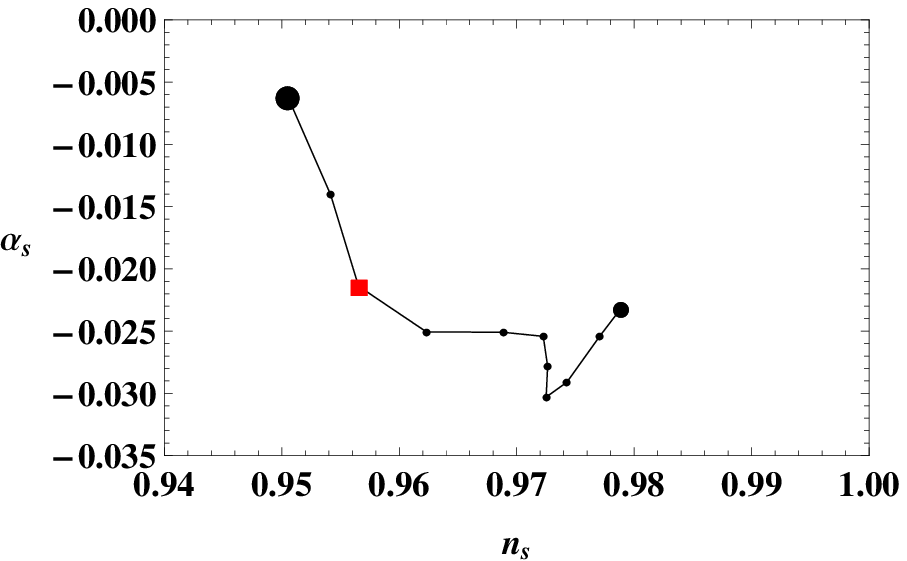}
        \end{center}
      \end{minipage}
\hspace{1mm}
      \begin{minipage}{0.32\hsize}
        \begin{center}
          \includegraphics[clip, width=\hsize]{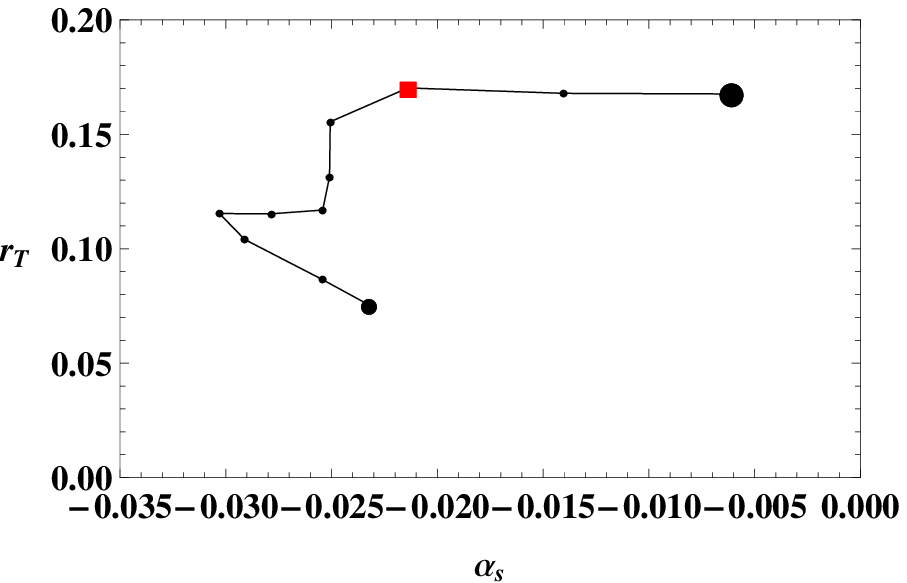}
        \end{center}
      \end{minipage}
  \end{center}
  \begin{center}
      \begin{minipage}{0.32\hsize}
        \begin{center}
          \includegraphics[clip, width=\hsize]{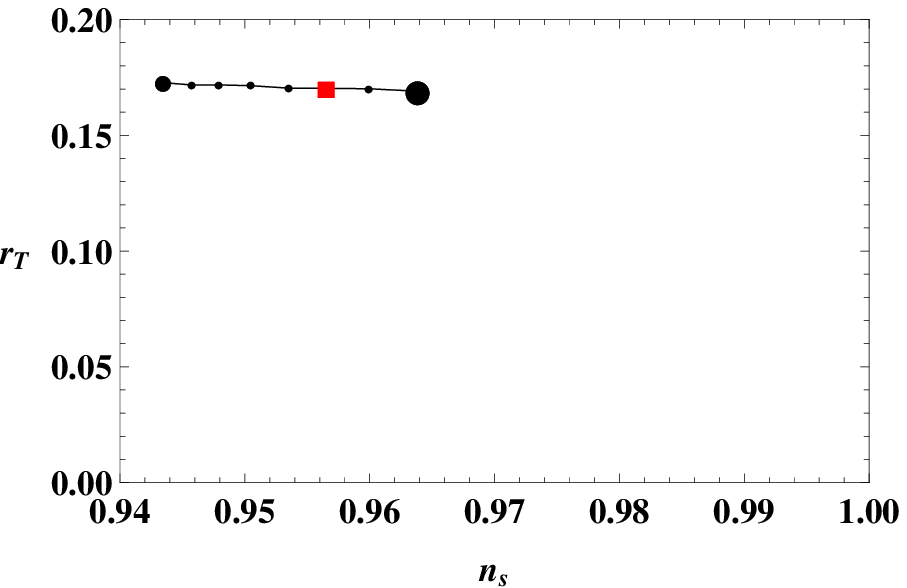}
        \end{center}
      \end{minipage}
\hspace{1mm}
      \begin{minipage}{0.32\hsize}
        \begin{center}
          \includegraphics[clip, width=\hsize]{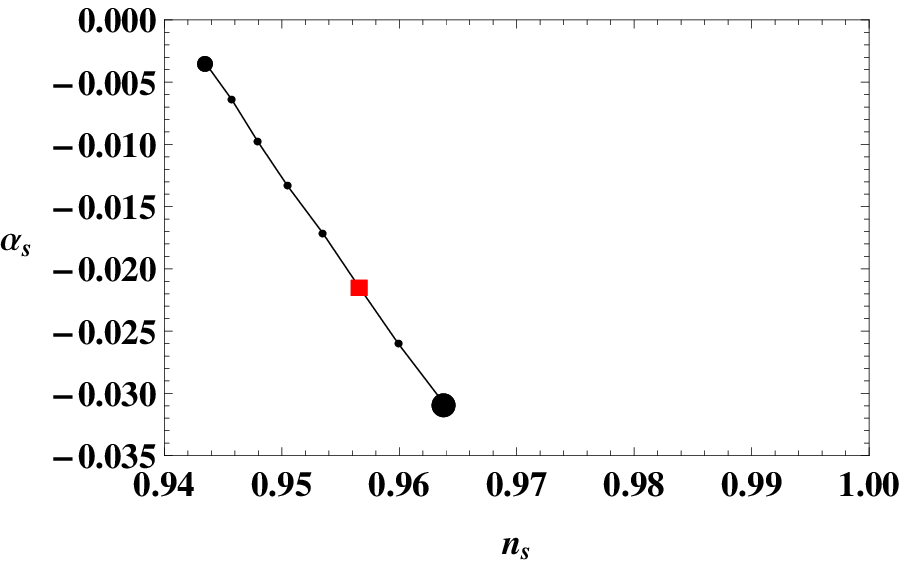}
        \end{center}
      \end{minipage}
\hspace{1mm}
      \begin{minipage}{0.32\hsize}
        \begin{center}
          \includegraphics[clip, width=\hsize]{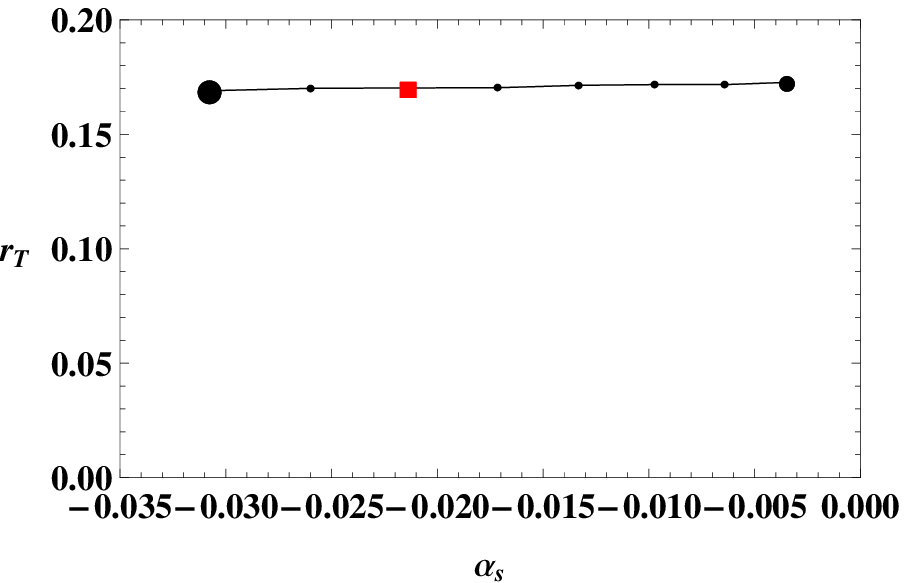}
        \end{center}
      \end{minipage}
  \end{center}
	\caption{
Plots in $(n_s,~r_T)$, 
$(n_s,~\alpha_s)$ or $(\alpha_s,~r_T)$-planes with varying $x$ (top) and with varying $x'$ (bottom).
			The red square corresponds to the parameters given by Table \ref{example}.
			From the black smaller circle to larger one,
			the dots run from $x=0.05$ to $0.55$ ($x'=0.04$ to $0.054$) at equal interval 0.05 (0.002).}
\label{x_dep}
\end{figure}
%
%
\begin{figure}[t]
  \begin{center}
      \begin{minipage}{0.32\hsize}
        \begin{center}
          \includegraphics[clip, width=\hsize]{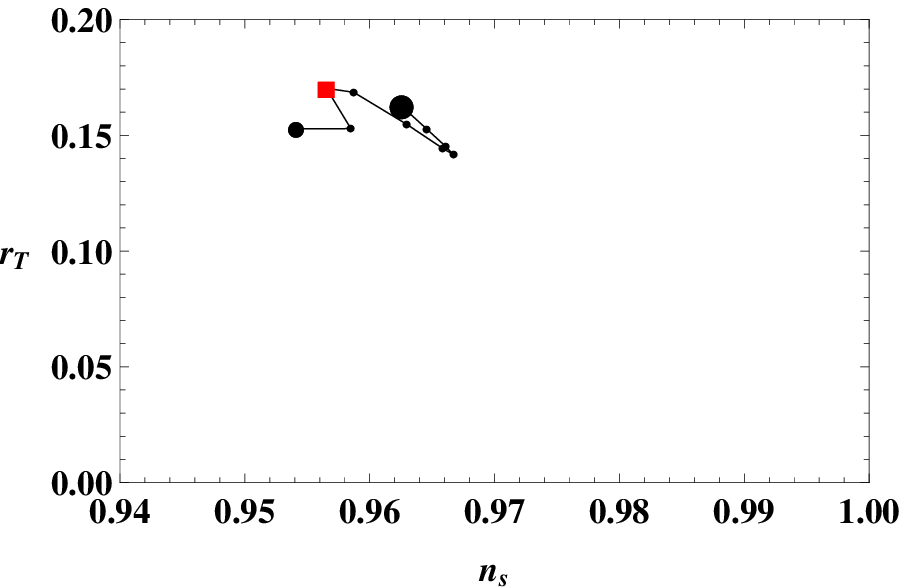}
        \end{center}
      \end{minipage}
\hspace{1mm}
      \begin{minipage}{0.32\hsize}
        \begin{center}
          \includegraphics[clip, width=\hsize]{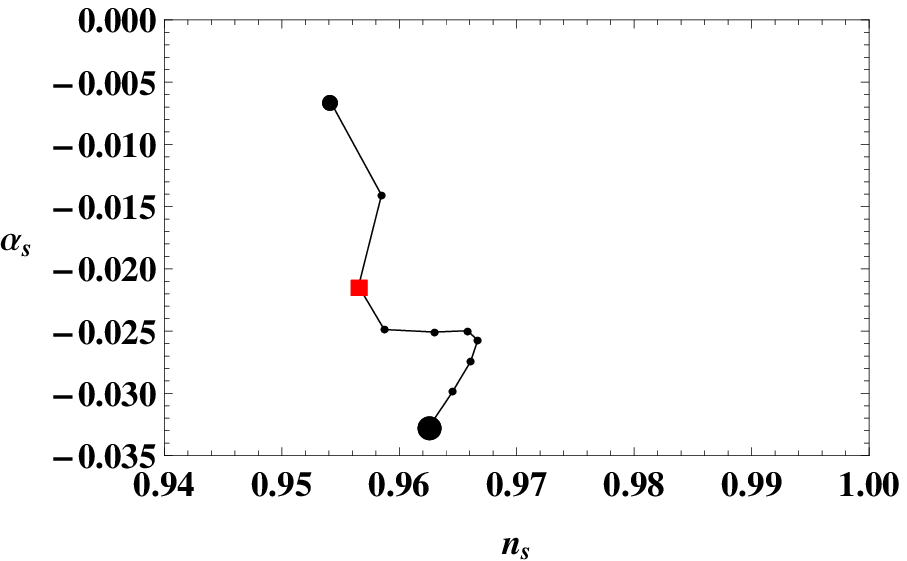}
        \end{center}
      \end{minipage}
\hspace{1mm}
      \begin{minipage}{0.32\hsize}
        \begin{center}
          \includegraphics[clip, width=\hsize]{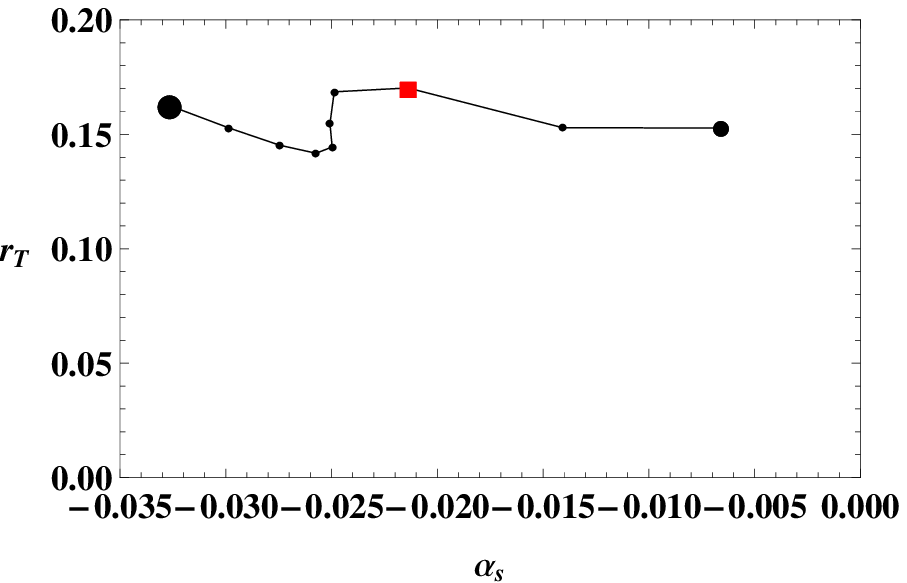}
        \end{center}
      \end{minipage}
  \end{center}
  \begin{center}
      \begin{minipage}{0.32\hsize}
        \begin{center}
          \includegraphics[clip, width=\hsize]{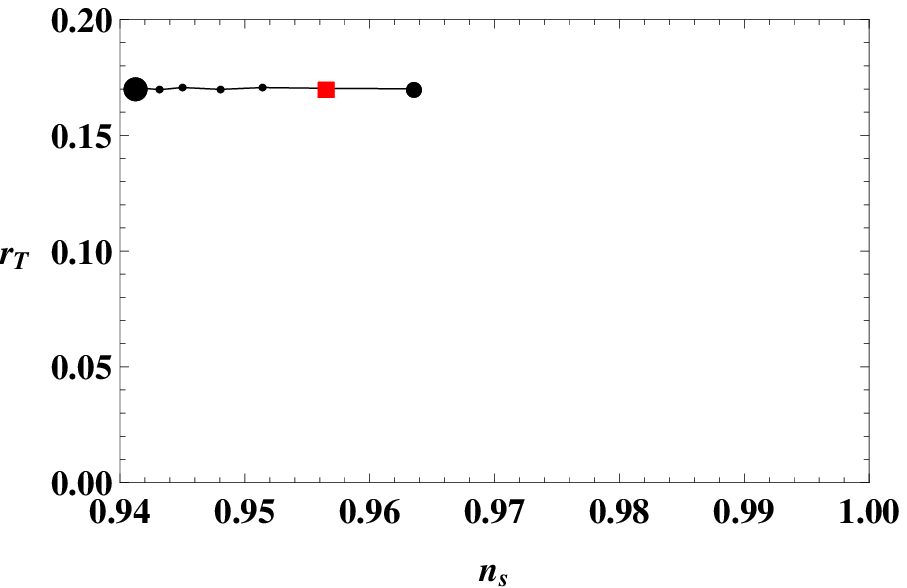}
        \end{center}
      \end{minipage}
\hspace{1mm}
      \begin{minipage}{0.32\hsize}
        \begin{center}
          \includegraphics[clip, width=\hsize]{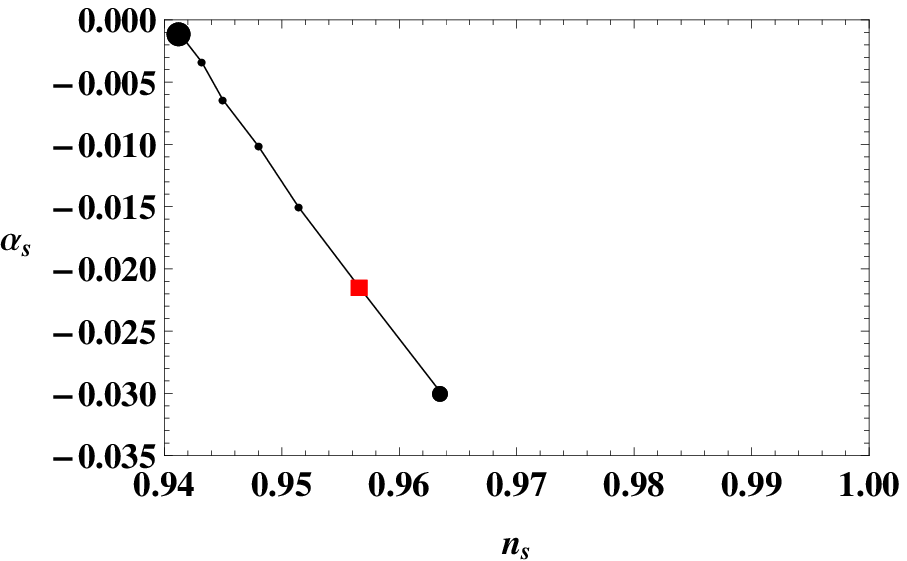}
        \end{center}
      \end{minipage}
\hspace{1mm}
      \begin{minipage}{0.32\hsize}
        \begin{center}
          \includegraphics[clip, width=\hsize]{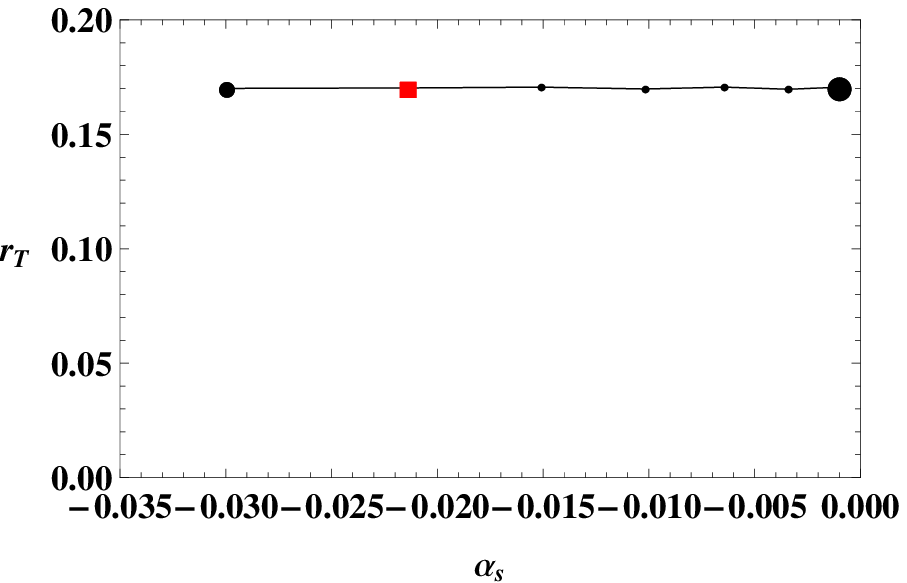}
        \end{center}
      \end{minipage}
  \end{center}
	\caption{
	Similar plots as in Fig.\,\ref{x_dep}, but with varying $A$ (top) and with varying $A'$ (bottom). 
			The red square corresponds to the parameters given by Table \ref{example}.
			From the black smaller circle to larger one,
			the dots run from $A=0.3$ to $0.75$ ($A'=0.009$ to $0.015$) at equal interval 0.05 (0.001).}
\label{A_dep}
\end{figure}
%
The parameters satisfy all Eqs. (\ref{theta}), (\ref{xA}), (\ref{rt_condition}) and (\ref{al_condition});
particularly, $(1-x \sin \theta ) \simeq 1.44$ and $x' /A'\, |\tan \theta'| \simeq 13.7$.
In the Fig.\,\ref{example_fig1},  we see that variation of $N$ becomes gentler when the slope of the potential becomes flatter, and that vice versa. Note that the sinusoidal functions do not give significant contributions to N.
%
Therefore, the discussion of previous subsection is valid;
$N$ can be estimated as in the case that there exists only the linear term in the potential, 
even though we have additional sinusoidal function corrections.


Next, we mention the parameter dependences of $n_s$, $r_T$ and $\alpha_s$.
In the following, we will use the inflaton value fixed at $\phi=10$.
First, we show $x$ and $x'$ dependences in Fig.\,\ref{x_dep}.
Since $\varepsilon \sim \eta \sim 0.01$ in the present parameter regions,
$\alpha_s$ is almost given by $\xi$.
The $\alpha_s$ depends on $x$ and $x'$ as in Eq.(\ref{xi2}), in which the denominator is given just by $\phi$. 
Thus, we can focus just on the numerator to understand the behavior.   
%
%
We see that $|\alpha_s|$ becomes larger for larger $x'$ while $|\alpha_s|$ becomes smaller for larger $x$,
 when we focus on the case that $\xi >0$.
%
%
As for the spectral index $n_s$,
we can obtain larger $n_s$ 
as $|\alpha_s|$ becomes larger owing to the relation in Eq.(\ref{ns_high2}).  
%
Note that 
now $\phi = 10$ is fixed,
the parameter $\delta ~(= \theta - \phi x/A)$
varies as $x$ moves to an another value.
%
%
This is similar to dependences on $x'$.
If we neglected the constant term $v_0$ in the potential,
$r_T$ would be monotonously larger as $x$ is large.
However, this is not the case: the zigzag lines in Fig.\,\ref{x_dep} suggests that the constant term cannot be neglected in some cases.

In Fig.\,\ref{A_dep}, we show similar plots with varying $A$ and $A'$ as in Fig.\,\ref{x_dep}.
The behavior is very similar, but opposite to those of $x$ and $x'$ in the parameter space
when one changes them from smaller ones to bigger ones.
This is because combinations of $x/A~(= 1/f)$ and $x'/A' ~(= 1/f')$ 
are important in the observable parameters.
The changes of $A$ and $A'$ do not 
give significant contributions to $r_T$, 
which is almost determined by $x \sin\theta$ (see Eq.(\ref{appro2})).
%
%
The zigzag lines are also due to the constant term in the potential.

\begin{figure}[t]
  \begin{center}
      \begin{minipage}{0.32\hsize}
        \begin{center}
          \includegraphics[clip, width=\hsize]{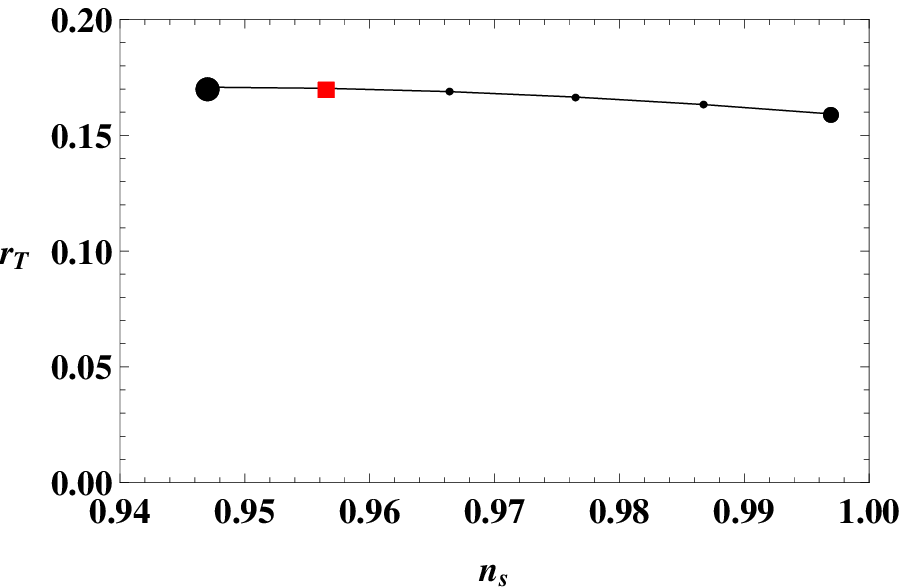}
        \end{center}
      \end{minipage}
\hspace{1mm}
      \begin{minipage}{0.32\hsize}
        \begin{center}
          \includegraphics[clip, width=\hsize]{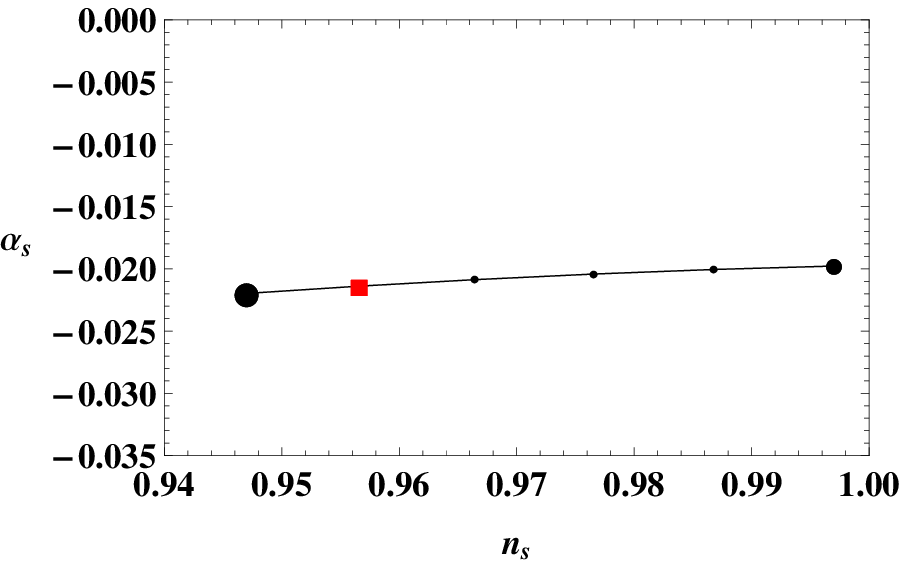}
        \end{center}
      \end{minipage}
\hspace{1mm}
      \begin{minipage}{0.32\hsize}
        \begin{center}
          \includegraphics[clip, width=\hsize]{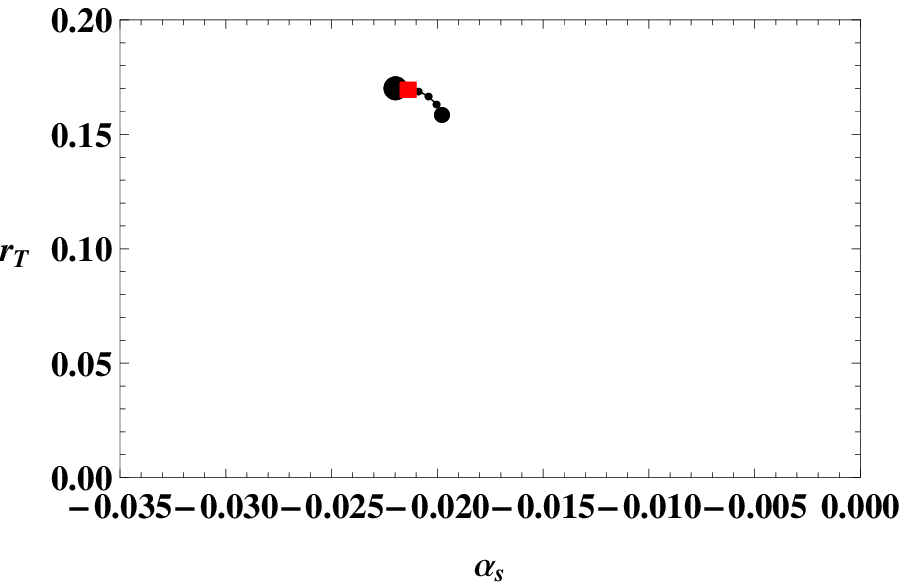}
        \end{center}
      \end{minipage}
  \end{center}
  \begin{center}
      \begin{minipage}{0.32\hsize}
        \begin{center}
          \includegraphics[clip, width=\hsize]{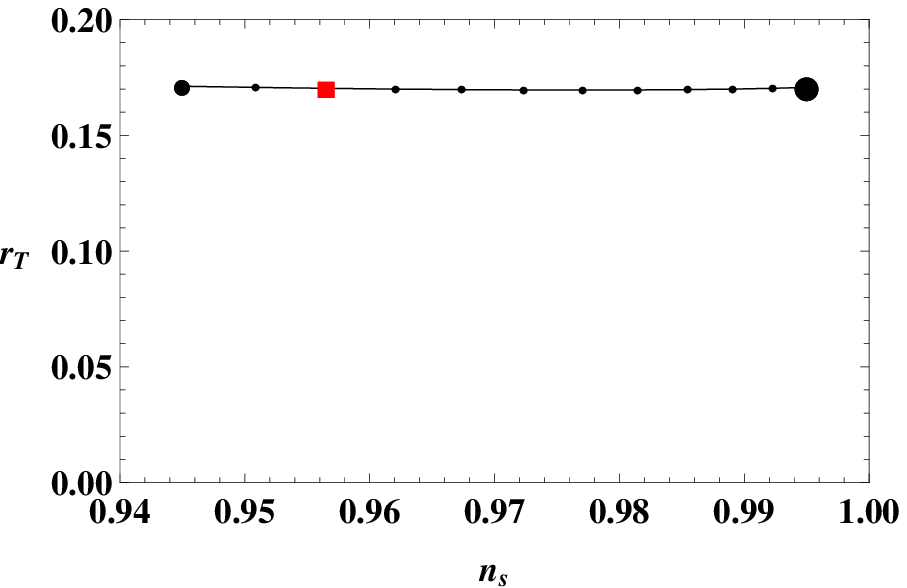}
        \end{center}
      \end{minipage}
\hspace{1mm}
      \begin{minipage}{0.32\hsize}
        \begin{center}
          \includegraphics[clip, width=\hsize]{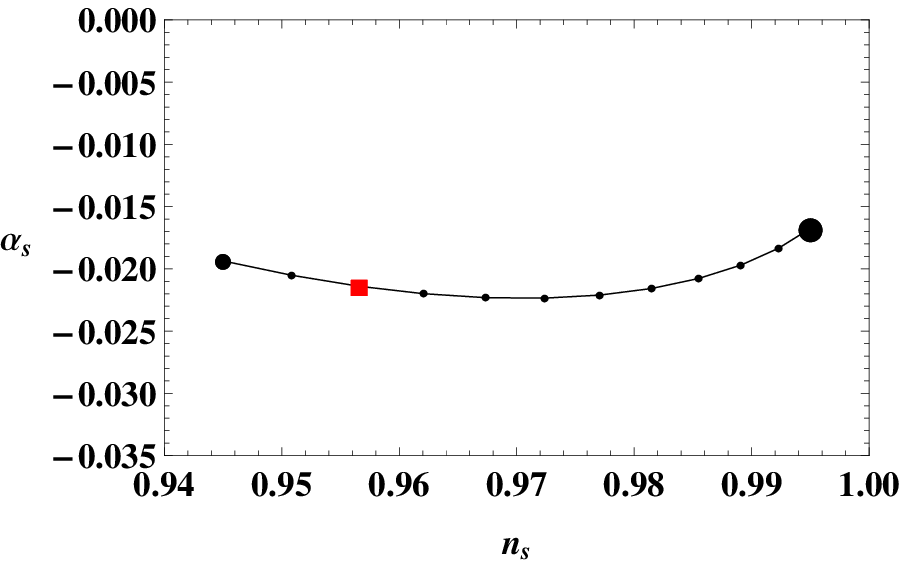}
        \end{center}
      \end{minipage}
\hspace{1mm}
      \begin{minipage}{0.32\hsize}
        \begin{center}
          \includegraphics[clip, width=\hsize]{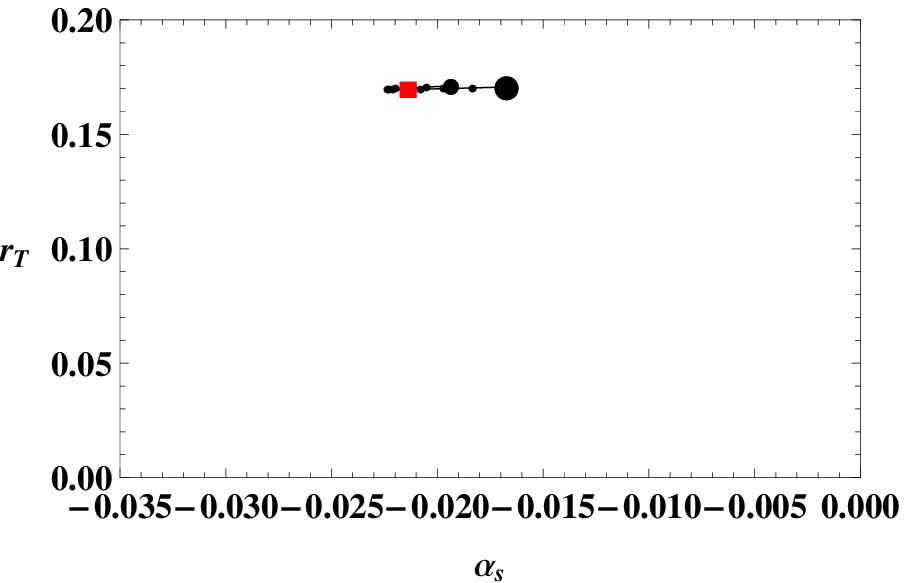}
        \end{center}
      \end{minipage}
  \end{center}
	\caption{
Similar plots as in Figs.\,\ref{x_dep} and Fig.\,\ref{A_dep}, but with varying $\theta$ (top) and with varying $\theta'$ (bottom).
			The red square corresponds to the parameters given by Table \ref{example}.
			From the black smaller circle to larger one,
			the dots run from $\theta=240 \pi/180$ to $265 \pi/180$ ($\theta'=60 \pi/180$ to $115 \pi/180$) at equal interval $5 \pi/180$ ($5 \pi/180$).}
\label{theta_dep}
\end{figure}

Finally, we show $\theta$ and $\theta'$ dependences in Fig.\,\ref{theta_dep}.
%
%
We can see that $n_s$ is significantly changed as they vary, while $r_T$ and $\alpha_s$ are not so.
Within the parameter range of $235\pi/180 \leq \theta \leq 265\pi/180$, 
one finds that $n_s$ can decrease rapidly as $\theta$ becomes large.
This is because
we obtain large $\sin \theta$ (and small $\cos \theta$), and hence
$\varepsilon$ becomes larger while $\eta$ becomes smaller.
Within the range of $50\pi/180 \leq \theta' \leq 115\pi/180$, on the other hand, $n_s$ becomes larger
since $\eta$ gets increased as $\theta'$ becomes large.
This is because
$\cos \theta'$ has a great effect on $\eta$ because of $x/A < x'/A'$, while $\sin \theta'$ is not so against 
$\varepsilon$ because of $x>x'$. 
%

As for the decay constants,
we are considering sub-Planckian ones.
%
As an example, we have taken $f = 0.89$ and $f' = 0.2$ for the parameters in Table \ref{example}.
We show other examples with the smaller decay constants in Table \ref{example2}.
To realize the favored value of $r_T$, $x$ should satisfy a restriction $0.4 \lesssim x \lesssim 0.6$.
For given parameters of $x$ and $A$ (or $f$),
$x'$ and $A'$ (or $f'$) should be much smaller than them to realize observable quantities on which we are focusing.
%
Note that $\theta$ and $\theta'$ can be used to fit $n_s$.
It is also viable that the decay constants much smaller values than the above examples for our purpose.
%
For instance, we can take $f = O(10^{-3})$ and $f' = O(10^{-4})$.
However, we do not show such examples,
 because we need fine-tuning to realize such small decay constants.
Nevertheless, it is worth noting that
 the inflaton potential (\ref{potential}) can realize favored values of all $n_s$, $r_T$ and $\alpha_s$
 with sub-Planckian decay constants.
%
\begin{table}[t]
\begin{tabular}{|cccc|cccc|cccc|}\hline
~~~$x$~~~ & ~~~$A$~~~ & ~~~$f$~~~ & ~~~$\theta$~~~ & ~~~$x'$~~~ & ~~~$A'$~~~ & ~~~$f'$~~~ & ~~~$\theta'$~~~
& ~~~$N$~~~ & ~~$n_s$~~ & ~~~$r_T$~~~ & ~~~$\alpha_s$~~~ \\
\hline \hline
0.4 & 0.3 & 0.75 & ~$260 \pi / 180$~ & 0.05 & 0.008 & 0.16 & ~$60 \pi / 180$~ & 57 & 0.0962 & 0.14 & -0.028 \\	
0.4 & 0.2 & 0.5 & ~$250 \pi / 180$~ & 0.04 & 0.004 & 0.1 & ~$35 \pi / 180$~ & 57 & 0.0958 & 0.14 & -0.024 \\	
0.4 & 0.2 & 0.5 & ~$260 \pi / 180$~ & 0.01 & 0.0005 & 0.05 & ~$35 \pi / 180$~ & 56 & 0.0960 & 0.15 & -0.022 \\	
0.4 & 0.1 & 0.25 & ~$265 \pi / 180$~ & 0.02 & 0.001 & 0.05 & ~$65 \pi / 180$~ & 55 & 0.0963 & 0.15 & -0.026 \\	
\hline
0.5 & 0.4 & 0.8 & ~$250 \pi / 180$~ & 0.06 & 0.01 & 0.167 &~$50 \pi / 180$~ & 60 & 0.0963 & 0.16 & -0.029 \\ 
0.5 & 0.25 & 0.5 & ~$265 \pi / 180$~ & 0.03 & 0.003 & 0.1 & ~$70 \pi / 180$~ & 60 & 0.0959 & 0.17 & -0.026 \\ 
0.5 & 0.25 & 0.5 & ~$265 \pi / 180$~ & 0.008 & 0.004 & 0.05 & ~$60 \pi / 180$~ & 60 & 0.0961 & 0.17 & -0.025 \\ 
0.5 & 0.15 & 0.3 & ~$265 \pi / 180$~ & 0.03 & 0.002 & 0.067 & ~$70 \pi / 180$~ & 59 & 0.0960 & 0.17 & -0.026 \\ 
\hline
0.6 & 0.5 & 0.83 & ~$250 \pi / 180$~ & 0.06 & 0.01 & 0.167 & ~$50 \pi / 180$~ & 63 & 0.0959 & 0.19 & -0.030 \\ 
0.6 & 0.3 & 0.5 & ~$255 \pi / 180$~ & 0.04 & 0.004 & 0.1 & ~$50 \pi / 180$~ & 66 & 0.0965 & 0.18 & -0.024 \\ 
0.6 & 0.3 & 0.5 & ~$265 \pi / 180$~ & 0.01 & 0.0005 & 0.05 & ~$55 \pi / 180$~ & 66 & 0.0954 & 0.19 & -0.030 \\ 
0.6 & 0.2 & 0.33 & ~$255 \pi / 180$~ & 0.05 & 0.004 & 0.08 & ~$50 \pi / 180$~ & 60 & 0.0965 & 0.20 & -0.027 \\ 
\hline
\end{tabular}
\caption{Parameter sets which can realize the favored values of all $n_s$, $r_T$ and $\alpha_s$}
\label{example2}
\end{table}
%

We use the parameter region such that the slow roll conditions are not 
violated during inflation.
Otherwise, resonance effects may enhance the scalar spectrum and 
non-Gaussianity \cite{Chen:2006xjb,Flauger:2009ab}.

\section{Conclusion and discussion}

We have studied the axion monodromy inflation with two
sinusoidal terms, which are generated by non-perturbative effects.
In our model, we can realize $n_s \sim 0.96$, $r_T \sim 0.16$, $\alpha_s \sim - (0.02-0.03)$ 
and $N \geq 50$ at the same time.
To realize these, we need a small hierarchy among the energy scales between parameters, 
$0.1 \lesssim a_2/a_1 \lesssim 1$ and $0.001 \lesssim a'_2/a_1 \lesssim 0.1$, 
although it is not a huge hierarchy, in particular for $(a_2/a_1)^{1/4}$ and  $(a'_2/a_1)^{1/4}$.
Note that $a_2$ and $a'_2$ are parameters with dimension 4.
It is also remarkable that the decay constants in our model satisfy $f, f' < 1$, 
and we do not need super-Planckian decay constants.
In our case, the normalization of parameters is set for the inflation to be consistent with the PLANCK normalization: 
$a_1 \sim 10^{-10}$; $a_2 \sim 10^{-11} \gg a_2' \sim 10^{-13}$ with $f \sim f' \sim 0.1$.
Hence, the inflaton mass, $m_{\phi} \sim a_1^{1/2}$, is expected to be of order $10^{13}-10^{14}$\,GeV,
where we have expanded the potential $V_{\rm NS5}(\phi) = a_1 \sqrt{1+\phi^2}$ around the origin of $\phi$.\footnote{
Such a mass term is expected in the presence of anti-NS5-branes because of the supersymmetry breaking.
}

For axion inflation to remain natural,
 the axion must be periodic with some fundamental periodicity $f_\phi$.
In this case, we should have $f = f_\phi/n$ and $f' =f_\phi/n'$.
We have in general assumed for exploratory purposes that $f$ and $f'$ are independent,
 while the potential for $\phi$ is still generated by nonperturbative effects only.
Nonetheless, note that the second model of table II does satisfy this constraint
 with $(f_\phi, n, n') = (M_{\rm Pl}/2, 1, 5)$. 
The decay constants, $f$ and $f'$, of other models could also be realized
by $f_\phi \sim (0.5 - 1.0)\times M_{\rm Pl}$.
If $f_\phi$ is much smaller than $M_{\rm Pl}$ and the axion winds many times,
light modes would appear between D-branes \cite{Silverstein:2008sg,McAllister:2008hb,Kaloper:2008fb}.
Such light modes would affect the inflation potential, although $f_\phi$ in our model
would be large enough.
At any rate, it is important to discuss the origin of the decay constants $(f,f')$ in the potential. 
However, this is beyond our scope because a concrete string model is required through the moduli stabilization of the extra dimension.\footnote{Even if $f_\phi$ is not large enough and
such light modes appear, the Hubble-induced masses would be helpful to make such light modes massive
at least during the inflation and no effect on the inflation potential.
}

Our results have interesting implications in model building, in particular 
string cosmology.
Successful results require two non-perturbative effects with a small hierarchy 
as mentioned above.
If the monodromy inflation is embedded into the large volume scenario \cite{Balasubramanian:2005zx},
there exists long-lived modulus $\Phi$ in the bulk . Note that the Standard Model sector may be located near the NS5-brane
for successful reheating after the inflation through the axionic coupling between the inflaton and the gauge bosons
\cite{Grimm:2011dj}:\footnote{
It would be required that (more than) two axions exist near the NS5-brane or the Standard Model brane, i.e., $h^{1,1}_- \geq 2$,
because either of the axions is eaten by an anomalous $U(1)$ gauge boson.
} 
The decay width $\Gamma_{\phi} \sim m_{\phi}^3/f^2$ is obtained with the interaction 
$(\phi/f) F_{\mu}\tilde{F}^{\mu \nu}$. Note that $f \sim {\cal V}^{-1/3} \sim 0.1$.
The modulus can decay well after the inflation and reheat the universe finally: The decay width of the moduli is given by
$\Gamma_{\Phi} \sim m_{\Phi}^3$; $\Gamma_{\Phi}/\Gamma_{\phi} \sim f^2 \sim {\cal V}^{-2/3} \ll 1$ 
if $m_{\phi} \sim m_{\Phi}$.
Here, the modulus mass is expected to be rather larger than the Hubble scale during the inflation, 
i.e., larger than of order $10^{14}$\,GeV; the gravitino mass is also comparable to this scale. Otherwise, decompactification of the
extra dimension can take place \cite{Kallosh:2004yh}. Thus, $m_{\phi} \sim m_{\Phi} \sim 10^{14}$\,GeV can be justified.
The decay product from the modulus decay can contain almost massless axions, 
which can behave as the dark radiation \cite{Cicoli:2012aq}. 
The presence of such an extra relativistic state can ameliorate the tension between the result of the PLANCK and 
that of the BICEP2 for $n_s \gtrsim 0.96$ \cite{Dvorkin:2014lea}.
Further, the leptogenesis mechanism scenario can work for generating the right amount of the baryon asymmetry
if we have right-handed neutrinos \cite{Fukugita:1986hr}: $T_R \sim \sqrt{\Gamma_{\Phi}} \sim 10^{11}$\,GeV,
where $T_R$ is the reheating temperature obtained through the modulus decay.
As a consequence, we might lose a candidate of neutralino dark matter in the supersymmetric models then,
while the Standard Model Higgs mass can be consistent with the mass scale at $126$\,GeV \cite{Hebecker:2012qp}.
However, now there exist sectors generating non-perturbative effect,
so it is natural to expect hidden sectors on a D-brane other than the Standard Model sector. 
The hidden sector branes may wrap on the same homology class with the cycle wrapped by
the Standard Model branes to satisfy tadpole (consistency) conditions.
In this case, the feature of the hidden sector might become similar to the Standard Model sector; 
mirror dark matter in the hidden brane may explain the presence of dark matter after the modulus decay,
solving coincidence problems \cite{Higaki:2013vuv,Higaki:DPW}.


\section*{Acknowledgments}
This work was supported in part by the Grant-in-Aid for Scientific Research 
No.~25400252 (T.K.) and on Innovative Areas No.~26105514 (O.S.) and by Young Scientists (B) No.~25800169 (T.H.)
from the Ministry of Education, Culture, Sports, Science and Technology in Japan. 
%



\end{document}